\documentclass{aa}
\usepackage{graphics,supertabular}
\usepackage{longtable}
\usepackage{graphicx}
\usepackage{amsfonts}
\usepackage{txfonts}
\usepackage{amssymb}
\usepackage{xcolor}
\usepackage{soul}
\usepackage{float}

\begin{document}

\title{Multiwavelength and astrometric study of the DBS\,89$-$90$-$91 embedded clusters region}

\author{M.A. Corti\inst{1,2}, G.L. Baume\inst{1,3}, R.B. Orellana\inst{1,3} \& L.A. Suad\inst{4}} 

\authorrunning{Corti et al.}
\titlerunning{DBS89-90-91 embedded clusters region}

\institute{Facultad de Ciencias Astron\'omicas y Geof\'{\i}sicas (UNLP) \and 
  Instituto Argentino de Radioastronom\'ia (CONICET - CICPBA - UNLP) \and
  Instituto de Astrof\'isica La Plata (CONICET - UNLP) \and
  Instituto de Astronom\'ia y F\'isica del Espacio (UBA--CONICET), CABA, Argentina.}

   \date{Received \today; accepted \dots}

\abstract
{}
{Our main targets were to improve the understanding of the main properties of G316.8$-$0.1 (IRAS\,14416$-$5937) radio source where the DBS\,89$-$90$-$91 embedded clusters are located, to identify the stellar population present in this region, and to study the interaction of these stars with the interstellar medium.}
{We analyzed some characteristics of the G316.8$-$0.1 radio source consulting the SUMSS to study the radio continuum emission at 843 MHz and the H\,I SGPS at 21 cm. 
We also used photometric data at the $JHK$ bands in the region of DBS\,89$-$90$-$91 clusters obtained from the VVV survey and supplemented with 2MASS catalogue. The investigation of possible stars associated with the H\,II region was complemented with the astrometric analysis using the Gaia Early Data Release 3. To study the young stellar objects we consulted the mid-infrared photometric information from WISE, Spitzer$-$GLIMPSE Surveys, and MSX point source catalog.}
{The photometric and astrometric research carried out in the IRAS\,14416$-$5937 region allowed us to improve the knowledge about the DBS\,89$-$90$-$91 embedded clusters and their interaction with the interstellar medium. In the case of DBS\,89 cluster, we identified 9 astrophotometric candidate members and 19 photometric candidate members, whereas for DBS\,90$-$91 clusters we found 18 candidate photometric members. We obtained a distance value for DBS\,89 linked to G316.8$-$0.1 radio source of 2.9 $\pm$ 0.5 kpc. We also investigated 12 Class\,I YSOc, 35 Class\,II YSOc, 2 MYSOc and 1 CHII region distributed throughout the IRAS\,14416$-$5937 region. Our analysis revealed that the G316.8$-$0.1 radio source is optically thin at frequencies $\geq$ 0.56 GHz. The H\,II regions G316.8$-$0.1$-$A and G316.8$-$0.1$-$B have similar radii and ionized hydrogen masses of $\sim$ 0.5 pc and $\sim$ 35 M$_{\odot}$, respectively. The ionization parameter computed with the younger spectral types of adopted members of DBS\,89 and DBS\,90$-$91 clusters, shows that they are able to generate the H\,II regions. The flux density of G316.8$-$0.1$-$B H\,II region is lower than the flux density of G316.8$-$0.1$-$A H\,II region.}
{A photometric and astrometric research looking for the members of the DBS\,89-90-91 embedded clusters has been carried out. We could identify the earliest stars of the clusters as the main exciting sources of the G316.8-0.1 radio source and we have also estimated the main physical parameters of this source. We improve the knowledge of the stellar components present in Sagittarius-Carina arm of our Galaxy and its interaction with the interstellar medium.}

\keywords{stars: early-type $-$ proper motions $-$ catalogs $-$ stars: formation $-$ ISM: structure $-$ radio lines: ISM}

\maketitle

\begin{table*}
\begin{center}
\leavevmode
\caption{Region properties and their associated objects}            
\label{tab:regions} 
\begin{tabular}{lcccl}
\hline
\hline
& \multicolumn{3}{c}{Region properties} & \\
\hline
& Region A & Region B & Region C & Reference \\
\hline
$\alpha_{J2000}$  &  14:45:21.4 &  14:45:03.0 &  14:45:19.3 & This work \\
$\delta_{J2000}$  & -59:49:25.2 & -59:49:30.0 & -59:50:36.7 & This work \\
$Radius$ [']      &         1.3 &         1.0 &         0.8 & This work \\ 
\hline
\hline
& \multicolumn{3}{c}{Associated objects} & \\
\hline
Radio source      &   G316.8-0.1 -- A &   G316.8-0.1 -- B & -- & \citet{vig07} \\
IRAS source       & 14416$-$5937 -- A & 14416$-$5937 -- B & -- & \citet{vig07} \\
Embedded cluster  & DBS 90-91 & DBS 89 & DBS 90 (partial) & \cite{dut03} \\
Astrometric group & -- & group B & group C & This work (Sect. \ref{sec:astrometry}) \\
\hline
\end{tabular}
\end{center}
\end{table*}

\begin{figure*}
\centering
\includegraphics[trim=100 40 100 50, width =0.4\textwidth]{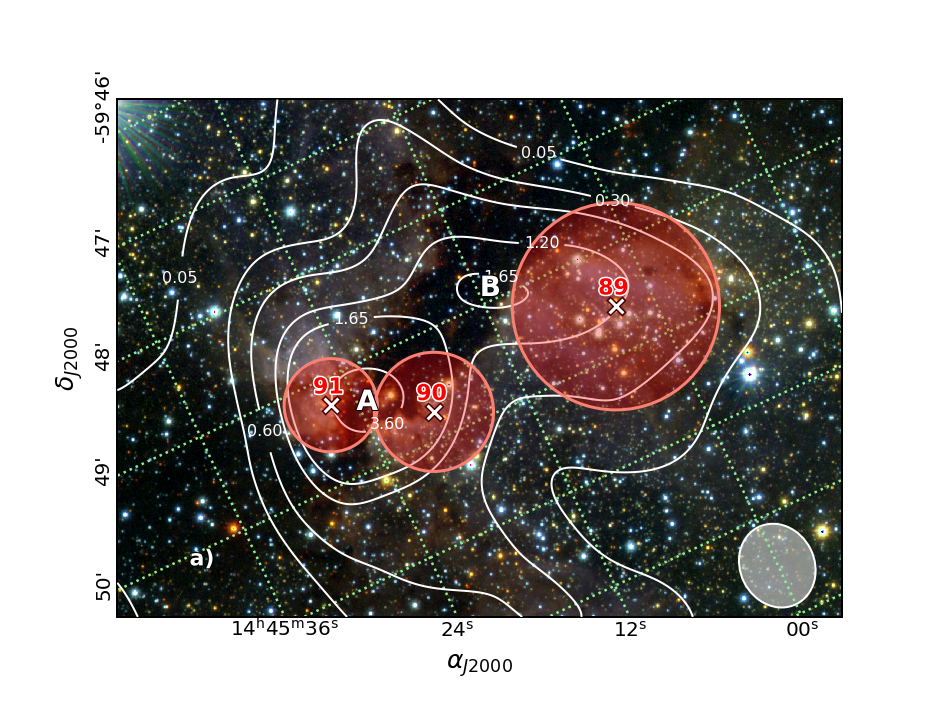}\hspace{0.15\columnwidth}
\includegraphics[trim=100 40 100 50, width =0.4\textwidth]{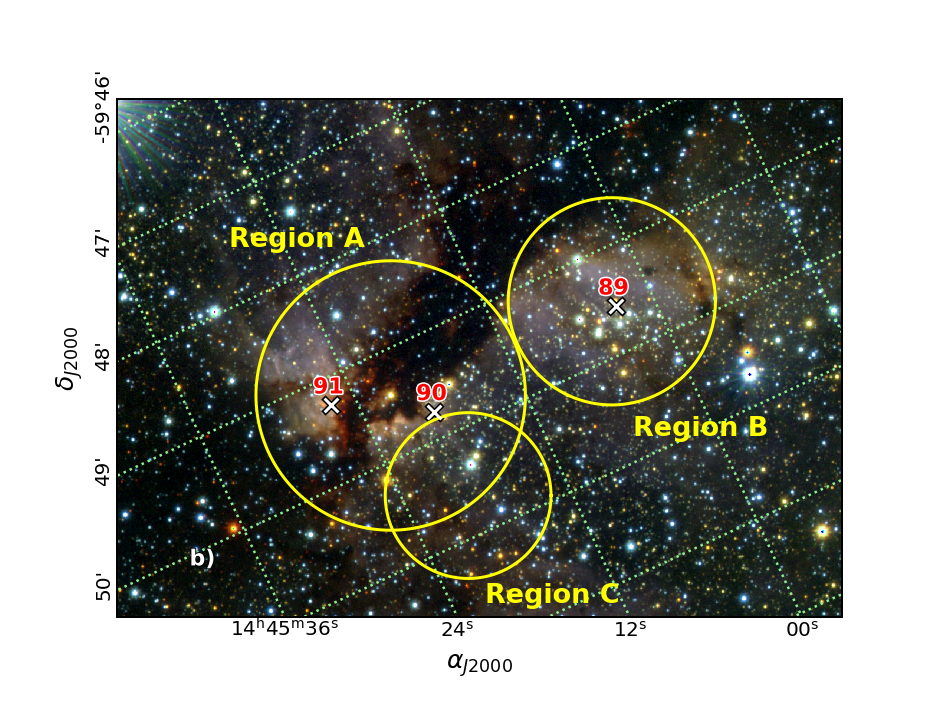}
\caption{$JHK$ false$-$color VVV image of \,7' $\times$ 5' size and centered at $\alpha_{J2000}$ = 14:45:12.2, $\delta_{J2000}$ = -59:49:27.6. a) White curves are radio continuum flux levels at 843 MHz (SUMSS) revealing the presence of two peaks "A" and "B" of G316.8$-$0.1 radio source and identified by \cite{vig07}. The white ellipse at the right-bottom corner indicates the corresponding radio beam size. Red circles represent the locations and mean sizes given by \cite{dut03} for embedded clusters DBS~89-90-91. {\bf b)} Yellow circles indicate the adopted "Region~A", "Region~B", and "Region~C" used along our work to study different stellar populations (see Sects.~\ref{sec:infrared} and \ref{sec:astrometry} for details). These regions were used to build the photometric diagrams showed in Fig.~\ref{fig:phot}. Centers for DBS~89-90-91 clusters are presented with white crosses.}

\label{fig:PhomIR}
\end{figure*}

\section{Introduction}
\label{Introd}

In recent years, technological advancements have made multiwavelength and homogeneous survey data for many star formation regions in the Galactic Plane.
Special cases are the Vista Variables in the V\'ia L\'actea (VVV; \citealt{min10}, \citealt{sai12}) survey and the Gaia Early Data Release 3 catalog ({\it Gaia} EDR3; \citealt{gaiacollaboration2020gaia}). The former has deeper and better spatial resolution images at near-infrared (IR) bands ($JHK$) than the Two Micron All Sky Survey (2MASS; \citealt{skr06}) ones, allowing to obtain photometric data of fainter objects with lower error values \citep{cor16}. On the other hand, $Gaia$ data provide the best astrometric measures at the moment (see reference \citealt{gaiacollaboration2020gaia} for details).

In particular, the fourth Galactic quadrant is a stellar {\it nursery} region where the G316.8$-$0.1 radio source is present. It is associated with the IRAS source 14416$-$5937 which nests the DBS\,89$-$90$-$91 embedded clusters \citep{dut03}. All of them revealed an active star-forming region and several studies have been performed over it up to date. Most of the works were focused on the study of the interstellar medium (ISM). For example \citet{cas87} analyzed the velocities of hydrogen recombination lines with 5~GHz observations and contributed to the knowledge of the spiral structure of our Galaxy. \citet{cas95} searched the Galactic methanol masers with 6.6~GHz observations and they concluded that this H\,II region has strongly variable features, and some weak features have been confirmed at velocities of $-$41 and $-$37 km s$^{-1}$ at several epochs. Later, \citet{bro96} identified a diversity of molecular lines using 98\,GHz observations, whereas \citet{bus06} found some maser sources analyzing 96\,GHz data. More recently, \citet{sam18} and \citet{dal18} studied G316.8$-$0.1 using multi$-$wavelength observations. In particular \citet{sam18} classified it as a bipolar H\,II region, whereas, \citet{dal18} compile an important number of masers detection. Subsequently, \citet{Wat19} analyzed the origin of the filamentary configuration present in the H\,II region and its influence on stellar feedback.

On the other hand, \cite{sha81} and \cite{vig07} carried out global studies including radio, optical and IR observations. In particular, \cite{sha81} provided one of the first approaches to the identification of the exciting source of the HII region, whereas \cite{vig07} analyzed photometric information of the bright ($J<15$) young star population, and the properties of the gas and dust in the region. Therefore, \cite{vig07} could obtain some parameters regarding the interaction between the star population and the ISM. \citet{vig07} also identified two peaks, named A and B, in both the IR observations of the IRAS source 14416$-$5937, and the radio data of G316.8$-$0.1. Additionally, it can be noticed that looking at the coordinates and sizes given by \cite{dut03}, DBS\,90$-$91 clusters are located around peak A, whereas DBS\,89 is located at the SW of peak B. (see Fig.~\ref{fig:PhomIR} and Table~\ref{tab:regions}).

Taking into account the recent information from the surveys indicated before, and since there are few known bipolar H\,II regions in the Galaxy, the investigation of G316.8$-$0.1 radio source zone with new data and taking a global view is relevant to better understanding the different processes involving young stars and their connection with the environment. In particular, \citet{sam18} indicate that they were not able to identify the exciting source of the G316.8$-$0.1. Therefore, we performed a deep and detailed photometric and astrometric study of the star populations in the zone of the DBS\,89$-$90$-$91 embedded clusters complemented with information of the surrounding ISM, and re-analyzed their fundamental parameters.

The structure of the paper is the follows. Observational data are presented in Sect. 2. Different surveys with radio data and photometric/astrometric information are briefly described. The analysis of our results is in Sect. 3. Finally, the discussion and conclusions are presented in Sect. 4 and 5, respectively.

\section{Observational Data}
\subsection{Photometric data}
\label{photom data}

We used photometric data of the objects located in the field of view (FOV) presented in Figs.~\ref{fig:PhomIR} and \ref{fig:SUMSSAyB}. We used the $JHK$ stacked images of the VVV survey and downloaded them from the VISTA Science Archive (VSA website\footnote{http://horus.roe.ac.uk/vsa/}). 
The selected images were obtained on April 3$^{rd}$, 2010 with an exposure time of 10 seconds in each band. The region under study has a high level of stellar concentration with a highly variable background in the IR. Then, we performed PSF photometry \citep{ste87} on the VSA images using {\sc IRAF}\footnote{{\sc IRAF} is distributed by NOAO, which is operated by AURA under a cooperative agreement with the NSF.} {\sc DAOPHOT} package to compute instrumental magnitudes. The obtained photometric tables were aperture-corrected for each filter to carry them to a final aperture size of 17 pixels in radius. Tables of different filters were combined using {\sc DAOMASTER} code (\citealt{ste92}). The calibration was carried out using the 2MASS catalog and the following transformation equations: \\

\indent $j_{inst} = j_0 + j_1 J + j_2~(J-H)$ \\
\indent $h_{inst} = h_0 + h_1 H + h_2~(H-K)$ \\
\indent $k_{inst} = k_0 + k_1 K + k_2~(H-K)$ \\

\noindent where $JHK$ are the 2MASS magnitudes, and ($jhk$)$_{inst}$ are the instrumental ones. The coefficients ($jhk$)$_x$ ($x = 0,1,2$) were computed using the {\sc FITPARAMS} task of {\sc IRAF PHOTCAL} package. The obtained values are shown in Table~\ref{tab:coefficients}.
We also used 2MASS catalog to complement the magnitudes of the bright stars ($J < 13.5$, $H < 12$, $K < 11$).

Additionally, we included the optical photometric data ($G$, $G_{BP}$, and $G_{RP}$ bands) given by the {\it Gaia} EDR3 covering the FOV. Since the studied objects can significantly change their fluxes between optical ($Gaia$) and IR bands, we did not consider photometric error constraints for optical bands, but a wide range of error ($e$ < 0.5) for near-IR bands. 

We also considered the mid-IR photometric information from the following catalogs: a) WISE \citep{cut14} at 3.4, 4.6, 12, and 22 $\mu$m bands ($W_1$, $W_2$, $W_3$, and $W_4$, respectively); b) Spitzer$-$GLIMPSE \citep{ben03} at 3.6, 4.5, 5.8, and 8.0 $\mu$m bands; and c) Midcourse Space Experiment \citep[MSX;][]{pri01} at 8.3, 12.1, 14.7, and 21.3 $\mu$m bands with a spatial resolution of $\sim$18". For WISE and GLIMPSE catalogs we selected sources with photometric uncertainty $<$ 0.2 mag in all bands and a relation signal-to-noise $>$ 7 for WISE sources. In the case of MSX point source catalog \citep{egan03}, sources were selected with variability and reliability flags zero and flux quality $Q$ > 1 in all bands.

We employed then the {\sc STILTS}\footnote{http://www.star.bris.ac.uk/$\sim$mbt/stilts/} tool to manipulate tables and to crossmatch the optical $Gaia$ data, the near-IR 2MASS/VVV data, and mid-IR WISE data in the FOV. In this procedure, we considered a maximum searching distance of 1" to match objects among 2MASS, VVV, and $Gaia$ catalogs and a distance of 4" for the WISE catalog. Additionally, to minimize possible mistaken associations between near-IR and mid-IR catalogs, we considered a difference lower than five magnitudes between the $K$ and $W_1$ bands. We obtained a catalog with photometric information for 3350 point objects in the FOV (see Fig.~\ref{fig:PhomIR}).

The photometric errors in the catalog were those provided by the {\sc DAOPHOT}, {\sc DAOMASTER} codes and the corresponding source catalogs.

\begin{table}
\centering
\caption{Calibration coefficients used for infrared observations together with the corresponding root-mean-square (rms) fit values.}
\label{tab:coefficients}
\begin{tabular}{ccccc}
\hline
      & $x = 0$          & $x = 1$          & $x = 2$          & rms  \\ 
\hline
$j_x$ & +1.49 $\pm$ 0.02 & +0.98 $\pm$ 0.01 & -0.02 $\pm$ 0.01 & 0.07 \\
$h_x$ & +1.48 $\pm$ 0.15 & +0.97 $\pm$ 0.01 & +0.06 $\pm$ 0.01 & 0.09 \\
$k_x$ & +1.62 $\pm$ 0.30 & +0.96 $\pm$ 0.02 & +0.10 $\pm$ 0.02 & 0.09 \\
\hline
\end{tabular}
\end{table}

\subsection{Radio data}
\begin{figure}
\centering
\includegraphics[trim=120 40 70 50, width =0.4\textwidth]{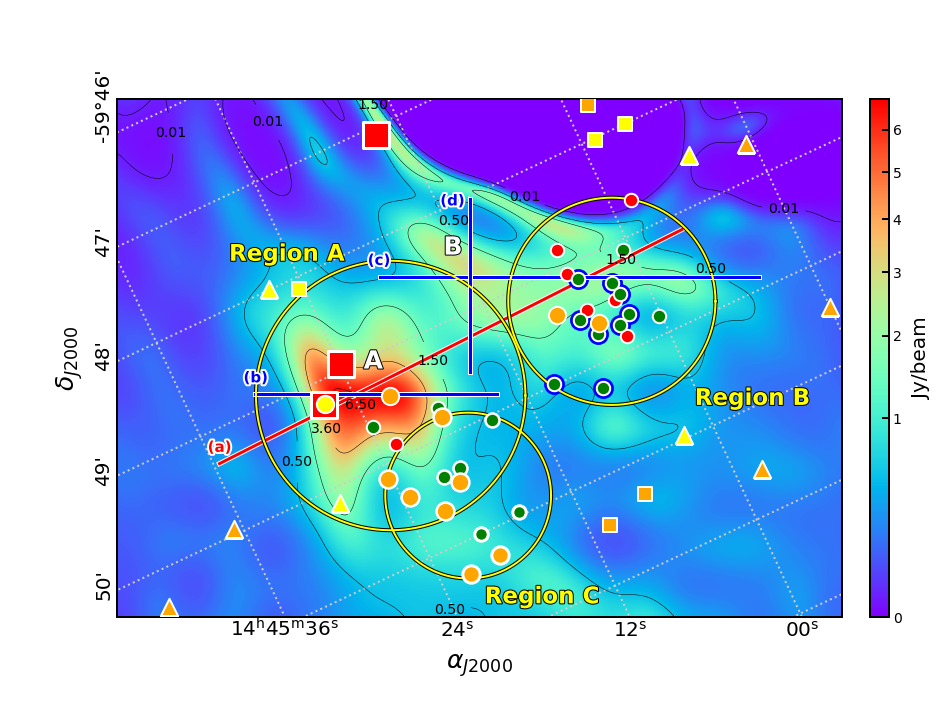}
\caption{Decovolved radio continuum image at 843 MHz (SUMSS) of the same size and center as Fig. \ref{fig:PhomIR}. Selected regions\,A, B, and  C (see Sect.\ref{sec:infrared}) are indicated by large yellow circles. Red and blue lines show the section, extension e inclination of the cuts made in different parts of the H\,II region. Symbols indicate the identified members in each studied region and identified YSOs in the surrounding. Green symbols correspond to MS stars and red ones to stars with probable IR excess. Yellow and orange symbols indicate Class I and II YSOs, respectively. Circles correspond to those identified using K band and WISE data, triangles those using only WISE data, and squares using GLIMPSE data. Big red squares indicate objects listed in Table \ref{YSOs3}. Green symbols with blue edges are the astrophotometric members presented in Table \ref{tab:parallax}.}
\label{fig:SUMSSAyB}
\end{figure}

\begin{figure}
\centering\includegraphics[width=0.45\textwidth]{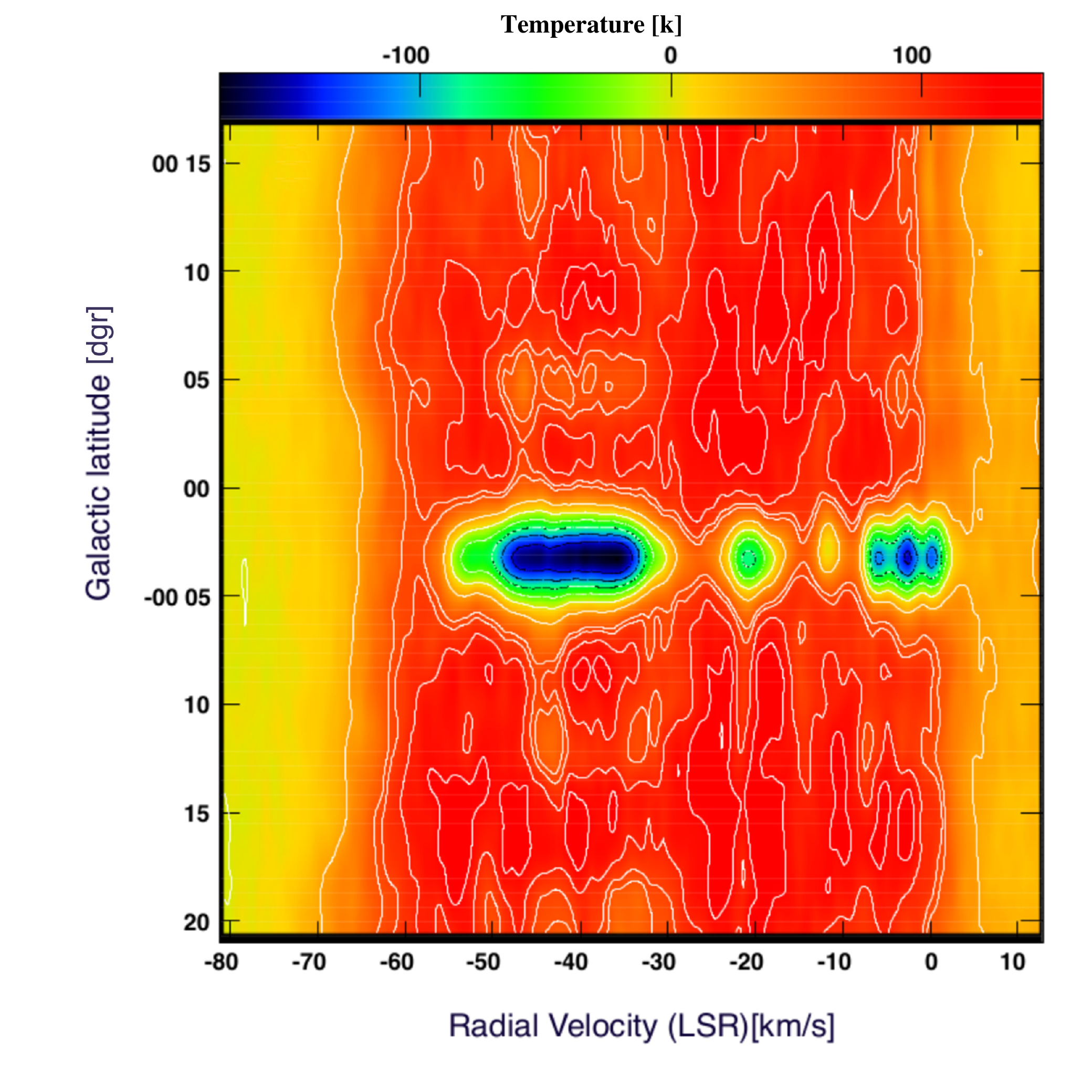}
\caption{Radio image at 21\,cm (SGPS) of H\,I emission distribution of the G316.8$-$0.1 radio source (l $\sim$ 317$^\circ$). H\,I absorption in the line of sight is indicated with blue and green colors. Contours have a spacing of 28~K being the first at $-$78~K. Colour bar shows the colors associated with different H\,I brightness temperature values in Kelvin units}.
\label{fig:Radio}
\end{figure}

We study some characteristics of the G316.8$-$0.1 radio source consulting suitable surveys. We employed the Sydney University Molonglo Sky Survey (SUMSS; \citealt{boc99}) that provides data with radio continuum emission at 843 MHz with a synthesized elliptical beam, 45"cosec |$\delta$| $\times$ 45". We used them to study the density flux intensity, the angular size of G316.8$-$0.1 radio source and, to confirm that in the environment close to this H\,II region there were no other radio H\,II regions (Fig. \ref{fig:SUMSSAyB}). The Southern Galactic Plane Survey (SGPS; \citealt{mcc05}) H\,I datacube provides data of the Galactic plane over the fourth quadrant with $2\farcm2$ angular resolution, $\Delta$V = 0.82~km~s$^{-1}$ velocity resolution, and 0.2 K rms noise brightness temperature ($T_b$). With this survey, we study the kinematic distance to G316.8$-$0.1 (Fig.~\ref{fig:Radio}).

\begin{figure}
\centering
\includegraphics[trim=70 70 70 70, width=\columnwidth]{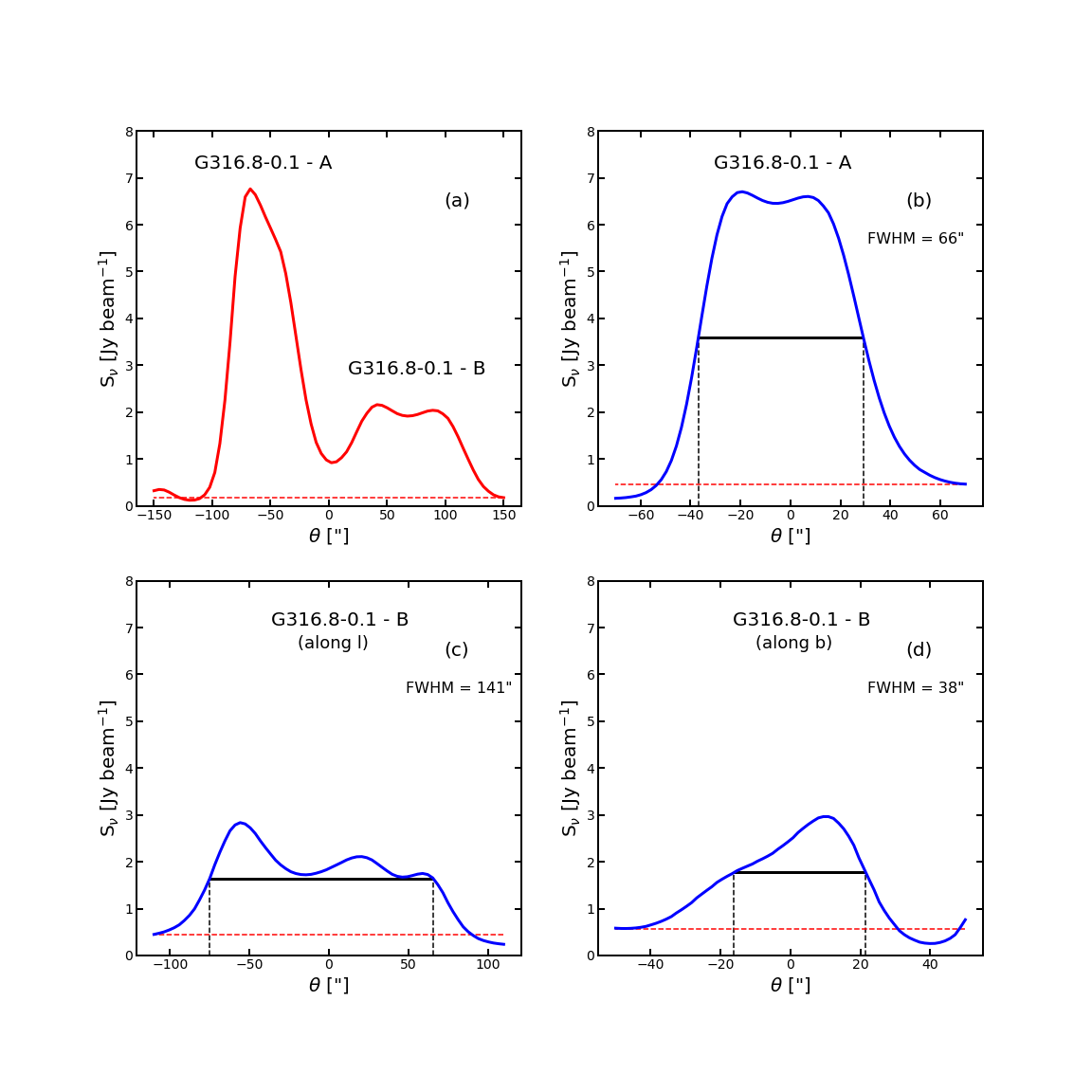}
\caption{H\,I profiles obtained with: (a) diagonal cut at l = 316$^{\circ}$.79, indicated with a red line in Fig. \ref{fig:SUMSSAyB}; (b) cut at l = 316$^{\circ}$.81, indicated with a blue line in Fig. \ref{fig:SUMSSAyB}. HPBW of G316.8$-$0.1$-$A profile is the deconvolved angular size of it; (c) cut at l = 316$^{\circ}$.77, indicated with a blue line in Fig. \ref{fig:SUMSSAyB}. HPBW of G316.8$-$0.1$-$B profile is the angular size of the part along the galactic plane; (d) cut at l = 316$^{\circ}$.79, indicated with a blue line in Fig. \ref{fig:SUMSSAyB}. HPBW of G316.8$-$0.1$-$B profile is the angular size of the part perpendicular to the galactic plane. Red slashed lines represent the adopted background levels.}
\label{fig:PerfilRB}
\end{figure}

\begin{figure}
\centering
\includegraphics[trim=30 20 30 40, width=0.45\textwidth]{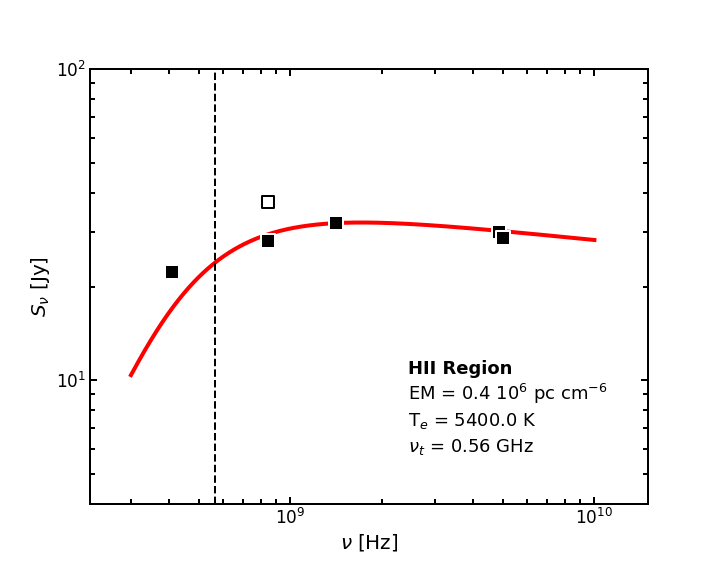}
\caption{Flux density measured at five different frequencies (see Sect. \ref{Analysis_Radio}) for all the G316.8$-$0.1 H\,II region. Red curve indicates the best-fitted radio spectrum of a thermal free$-$free emission model the indicated parameters. Only filled squares were used in the fits. Dashed vertical line shows the turnover frequency ($\nu_t$) value.}
\label{fig:SED}
\end{figure}

\subsection{Astrometric data}
\label{selection}

The astrometric analysis of the region was performed using the 
{\it Gaia} EDR3.
The reference epoch for {\it Gaia} EDR3 is 2016.0 and the catalog is essentially complete between $G$ = 12 and $G$ = 17 magnitude.

The astrometric data include the five astrometric parameters (position, parallax, and proper motion) for 1468 billion sources. 
The uncertainties in position, parallax, and proper motion according to the $G$ magnitude are given in Table~\ref{tabla_uncertainties} \citep{lin21}. The parallax zero point deduced from the extragalactic sources is about $-$17 $\mu$as \citep{lin21}. The probabilistic distances derived from the parallax are obtained from \cite{bai21}.

\begin{table}
\caption{ Typical uncertainties values for the 5-parameter solutions at different $Gaia$ magnitudes ($G$).}
\label{tabla_uncertainties} 
\centering                                      
\begin{tabular}{cccc} 
\hline
\hline 
 & $u_{Pos}$ & $u_{Plx}$ & $u_{pm}$ \\ 
 &  [mas] &  [mas] &  [mas\,yr$^{-1}$] \\
\hline
 $G$ < 15 & 0.01-0.02 &  0.02-0.03 & 0.02-0.03 \\
 $G$ = 17 &  0.05 &	 0.07 &  0.07 \\
 $G$ = 19 &   0.4 &	 0.5 &   0.5 \\
 $G$ = 21 &   1.0 &	  1.3 &  1.4 \\
\hline                                      
\end{tabular}
\parbox{7cm} {Note: Columns give the uncertainties for position ($u_{Pos}$), parallax ($u_{Plx}$), and proper motion ($u_{pm}$).}
\end{table}

We adopted a circular region with centre at ($\alpha_{J2000}$, $\delta_{J2000}$) = (14:45:12.0, -59:49:12.0) and 3.0’ in radius. Then, we selected 898 stars from {\it Gaia} EDR3 of which 801 
have the five astrometric parameters necessary for our analysis. To ensure the quality of the astrometric results, we make a selection of the stars taking into account the following criteria: 
i) the RUWE parameter ${\leq}$ 1.4 \citep{lin21} because this value is a quality indicator of the astrometric solution; ii) stars with Dup parameter equal to 1 are not selected because this value indicates probable  astrometric or photometric problems; iii) the values of the components of the proper motion are between -20 and +20 $mas\,yr^{-1}$ because a previous visual inspection of the Vector Point Diagram (VPD) of the region showed low proper motions for the possible embedded clusters; iv) stars without magnitude $G$ have not been included; 
 v)  systematic errors in the proper motion due to the angular scale of the studied region were also analyzed. In our sample, no star had a proper motion within uncertainties smaller than the systematic error given in Table 7 of \cite{lin21}. With these conditions, the sample is reduced to 754 sources.

\section{Analysis and Results}

\subsection{Interstellar medium}
\label{Analysis_Radio}

Following \citet{vig07}, G316.8$-$0.1 radio source main peaks, A and B, are associated with those of the IRAS source (see Table \ref{tab:regions}). Since SUMSS radio data have a relatively low spatial resolution, we carried out a deconvolution procedure over them to obtain a more detailed picture of the zone (Fig.~\ref{fig:SUMSSAyB}). The Richardson-Lucy method (\citealt{Richardson1972,Lucy1974})  was applied since it is able to preserve the flux information. Then, we obtained some characteristic profiles of the resulting image, choosing those including peaks A and B (see Figs.~\ref{fig:SUMSSAyB} and \ref{fig:PerfilRB}). The shape of G316.8$-$0.1$-$A is approximately circular with a deconvolved angular size of $\theta_l~=~\theta_b$~=~66", obtained from the measured half-power beam width (HPBW). The cut made to G316.8$-$0.1$-$A and shown in Fig.~\ref{fig:PerfilRB}b has a double peak (see Sect.~\ref {sec:p_objs} for details). The shape of G316.8$-$0.1$-$B is irregular and something elongated along the Galactic plane. Its deconvolved HPBWs values at each axis are $\theta_l$ = 141" and $\theta_b$ = 38", whereas the  corresponding geometric mean is ($\theta_l$$\times$$\theta_b)^{1/2}\simeq$73".

The flux density at 843 MHz was estimated for both regions.
Using {\it tvstat} AIPS{\color {blue}\footnote{http://www.aips.nrao.edu/index.shtml}} task, we obtained the background average flux density, $S_{bg}$ = 2.6 $\times$ 10$^{-2}$ Jy\,beam$^{-1}$, from the mean flux density values at three different positions near these sources. The root mean square of $S_{bg}$ was {\it rms}$_{S_{bg}}$ = 5.2 $\times$ 10$^{-2}$ Jy\,beam$^{-1}$ and it was used to distinguish the signal from the noise. The lower isophote presented in Fig.~\ref{fig:SUMSSAyB} corresponds to 0.5 Jy\,beam$^{-1}$, more than 3 times the {\it rms}$_{S_{bg}}$ value. We obtained the density flux value of each source and then subtracted the same $S_{bg}$ value for both regions. The results were $S_{\nu RA}$ = 18 Jy and $S_{\nu RB}$ = 9 Jy for sources G316.8$-$0.1$-$A and B, respectively. The density flux value measured across  all H\,II region was $S_{\nu}$~=~28~Jy.

When an ionized gas cloud is examined over the entire frequency range, the received signal will be affected by opacity depending on the frequency of the energy. At high frequencies, the plasma must become optically thin and the spectrum of the observed radio emission will become approximately flat. This is the behavior observed with the flux density measured at three different frequencies (see Fig.~\ref{fig:SED}). 

The spectral index $\alpha$ ($S \sim \nu^{-\alpha}$) can be obtained from the measured flux density at two different frequencies, $\nu_1$ and $\nu_2$ ($\alpha = log(S_1 / S_2)/log({\nu_2}/{\nu_1})$). Using $\nu_1$ = 843 MHz, $S_{\nu 1}$ = 28 Jy and $\nu_2$ = 1415 MHz, $S_{\nu 2}$ = 32 Jy \citep{sha81}, we obtained $\alpha$ = $-$0.26. We estimated another value of spectral index, $\alpha$ = 0.05, using $\nu_1$ = 1415 MHz, $S_{\nu 1}$ = 32 Jy and $\nu_2$ = 4850 MHz, $S_{\nu 2}$ = 30 Jy \citep{kuch97}. The value of $S_{\nu}$ = 37.5 Jy for $\nu$ = 843 MHz \citep{vig07} was discarded because 
is higher than our estimation probably due to it was obtained in a larger region ($\sim$ 30 arcmin$^2$) and does not correspond to the study region of this work. With the goal to obtained the best adjust of the all flux density shown in Fig \ref{fig:SED} we incorporated two additional values, $S_{\nu}$ = 22.3 Jy at $\nu_2$ = 408 MHz and $S_{\nu}$ = 28.7 Jy at $\nu_2$ = 5000 MHz, both results of the \citet{sha70} work. 

\begin{figure}
\centering
\includegraphics[width=0.45\textwidth]{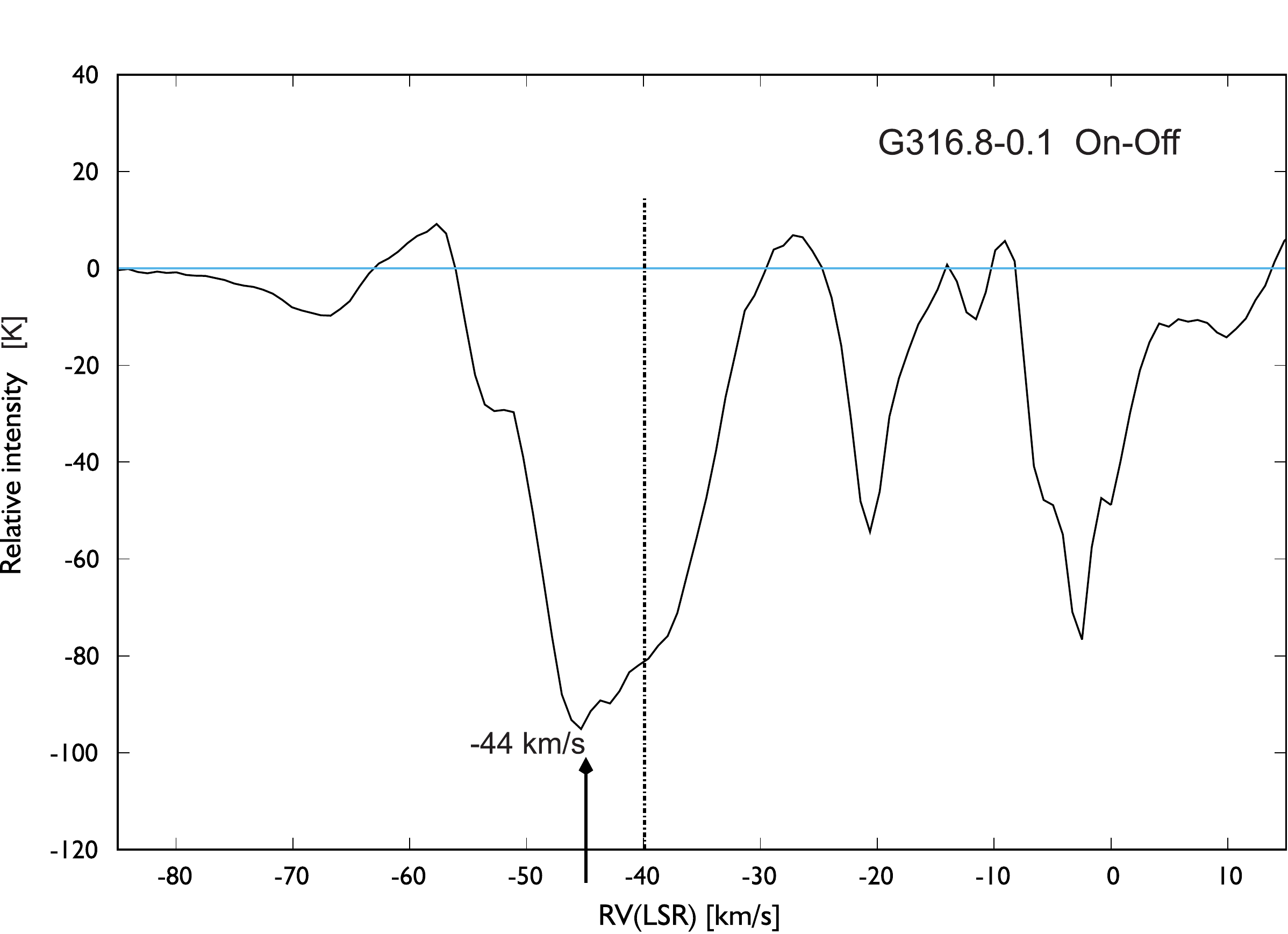}
\caption{Relative intensity of the G316.8$-$0.1 ("on$-$source") subtracted from the average of three sources of its environment ("off$-$source"). Higher intensity absorption would indicate the location of the H\,II region. Vertical line at $-$40 km s$^{-1}$ shows the velocity of $^{13}$CO (1$-$0) molecular line (emission) detected with the MSX survey by \citet{bus06}. Horizontal light blue line shows the zero value from which the absorption intensity is measured.}
\label{fig:both}
\end{figure}

We employed the 21 cm line emission map of the SGPS to study the H\,I absorption in the line of sight, which is indicated in Fig.~\ref{fig:Radio} with blue and green colors. This absorption could be a consequence of the temperature difference between the H\,II region and the distribution of the H\,I gas. The former has a continuum temperature ($T_c$) higher than the $T_b$ of the H\,I gas. In this case, A and B components of G316.8$-$0.1 radio source could not be angularly resolved therefore only a global study could be done, obtaining the same radial velocity measure for both. In order to determine the approximate radial velocity of this radio source, we obtained the “on” source profile in the brightest region of this using the H\,I data cube. Later, we obtained the “off” profile by doing the average profile with the ones obtained in three regions in the environment of the source. Assuming that the “on” and “off” spectra both sample the same gas, subtraction from one another removes the common emission. Fig.~\ref{fig:both} shows the “on”–“off” profile. The absorption feature observed in this figure at $-$44 km s$^{-1}$ Local Standard of Rest (LSR) radial velocity, corresponds to a kinematic distance of 3.3 $\pm$ 0.6 kpc for sources G316.8$-$0.1$-$A and B, according to the galactic rotation model of \citet{bra93}. These results are in agreement to that obtained by \citet{cas95}. The value of the radial velocity at $-$44~km~s$^{-1}$ matches pretty well with the $^{13}$CO (1$-$0) molecular line (emission, MOPRA telescope) and H\,I 21~cm data (absorption, SGPS survey) of the G316.8$-$0.1 obtained of the Red MSX survey \citep{bus06}. \citet{lon07} at $\sim$ 24~GHz with the Australia Telescope Compact Array (ATCA) measured the radial velocity, $V_{LSR}$, of the NH$_3$(1,1) and NH$_3$(2,2) molecular lines, with a spectral resolution of 0.2 and 12.7~km~s$^{-1}$, respectively, and NH$_3$(4,4) and NH$_3$(5,5) molecular lines, with spectral a resolution of 0.8~km~s$^{-1}$. The results obtained by them varies from $-$37.3 $\pm$ 0.3 to $-$42.2 $\pm$ 0.8 km s$^{1}$. 

To estimate other physical parameters of these regions we adopted the Gaussian model presented by \citet{mez67}. We computed the Stromgreen linear radius ($R_s$) as the deconvolved source angular size and the H\,II region distance of 2.9 kpc (see Sect. \ref{sec:dist}).

Additionally, we obtained the electronic density ($N_e$), the total mass of ionized hydrogen ($M_{HII}$) and the emission measure ($EM$) using equations and numerical values of each factor published by \citet{mez67}. We assumed, for both regions, the same  electronic temperature ($T_e$) value of 5400 K, obtained by \citet{cas95}. The calculated values for all of these parameters are shown in Table \ref{tab:AboutDistance}.

We estimated the optical depth, $\tau_{\nu}$ using the integral of the absorption coefficient along the line of sight. For the radio domain, \citet{mez67} have given the following useful expression valid for frequencies smaller than 10 GHz:

\begin{equation}
\label{optical_depth}
\tau_{\nu} \sim 8.235 \times 10^{-2}\, T_e^{-1.35}\, \nu^{-2.1} \int N_e^2\, dl.
\end{equation}   

The integral of the square of the electron density along the line of sight is the $EM$. We computed the frequency at which $\tau$ = 1, known as turnover, employing for this the $EM$ value 0.4 $\times$ 10$^6$ pc cm$^{-6}$ corresponding to all the G316.8$-$0.1 radio source. A typical frequency of turnover for free$-$free emission is 1 GHz and we obtained a value of $\nu$ = 0.56 GHz. 

\begin{table*}
\begin{center}
\leavevmode
\caption{Physical parameters of radio sources G316.8$-$0.1$-$A and B}
\label{tab:AboutDistance}
\begin{tabular}{ccccccc}
\hline
\hline
 ID & $R_S$ & $M_{HII}$ & $N_e$ (10$^3$) & $EM$ (10$^6$) & $\mu$ & $\theta_{HIIR}$ \\
   & [pc] & [$M_{\odot}$] & [cm$^{-3}$] & [pc\,cm$^{-6}$] & [pc\,cm$^{-2}$] & ['] \\
\hline
G316.8$-$0.1$-$A &  0.43 $\pm$ 0.09 & 36 $\pm$ 9 & 2.1 $\pm$ 0.10 & 2.8 $\pm$ 0.4 & 71 $\pm$ 15 &  1.0 \\
G316.8$-$0.1$-$B & 0.50 $\pm$ 0.11 & 32 $\pm$ 8 & 1.2 $\pm$ 0.09 & 1.0 $\pm$ 0.2 &  57 $\pm$ 12 &  1.2 \\

\hline
\end{tabular}
\end{center}
\end{table*}

\subsection{Stellar population methods}

As is revealed by near IR images (see Fig.~\ref{fig:PhomIR}), embedded clusters DBS\,89$-$90$-$91 are characterized by a relevant H\,II background emission. An overdensity of bright stars is evident for DBS\,89. However, DBS\,90 and DBS\,91 are fainter and a high absorption dust lane (SDC G316.786$-$0.044; \citealt{Peretto2009}) is present between them.

To study the stellar population associated with these embedded clusters, we selected three circular regions identified as Region\,A, Region\,B and Region\,C. Regions\,A, and B were defined {\bf to} include most of the cluster probable members. Therefore, their location and sizes were obtained using an iterative procedure looking for star overdensities in the IR image (Fig.~\ref{fig:PhomIR}) and the maximum amount of cluster members after to apply our photometric method (see Sect. \ref{sec:infrared}). 
Region C was defined as a result of our astrometric analysis. 
This analysis takes into account that a stellar group must show an over-density  in the sky as well as  in the VPD.
In Sect. \ref{sec:astrometry} we show how we first detect the over-density in the VPD and then identify the over-density in the sky by visual inspection.
Final adopted centers and sizes for each region are presented in Table~\ref{tab:regions} and plotted on Fig.~\ref{fig:PhomIR}. It can be noticed that whereas Region A  includes G316.8$-$0.1$-$A radio source peak, Region B is located to the west of G316.8$-$0.1$-$B peak.

\subsubsection{Photometric analysis}
\label{sec:infrared}

\begin{figure*}
\centering
\includegraphics[trim=100 10 100 50, width=\textwidth]{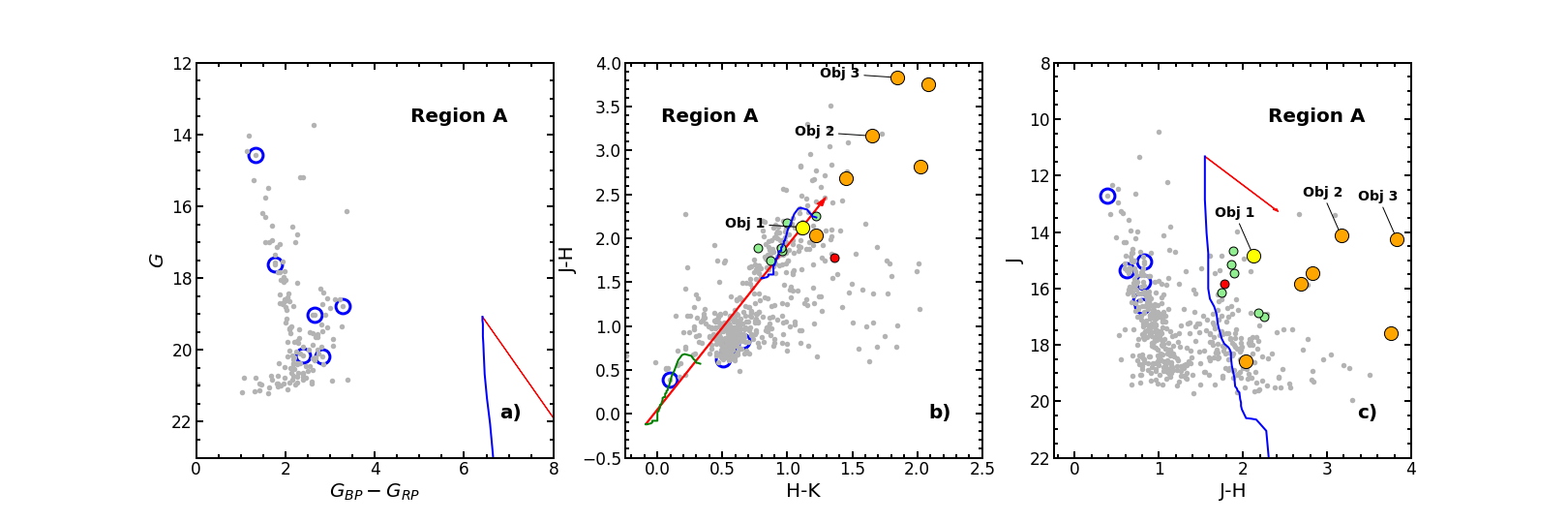}
\includegraphics[trim=100 10 100 50, width=\textwidth]{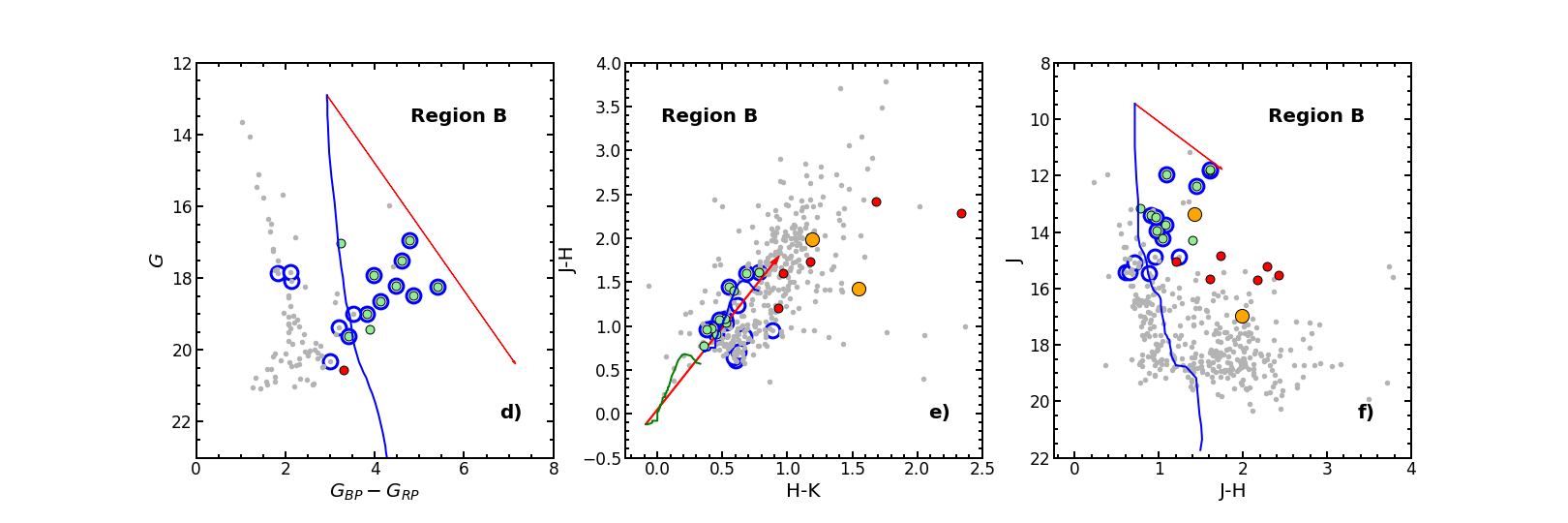}
\includegraphics[trim=100 10 100 50, width=\textwidth]{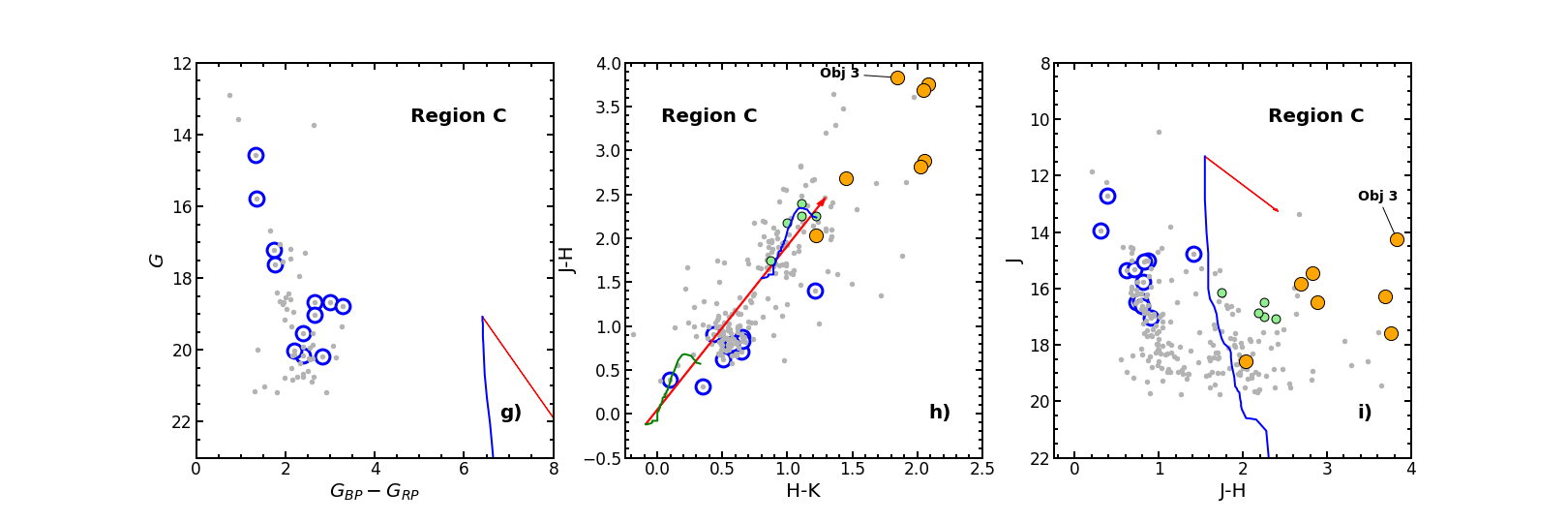}
\caption{$Gaia$ and IR photometric diagrams for Regions A, B and C. Symbols meaning as in Fig.~\ref{fig:SUMSSAyB}, but in this case hollow blue circles correspond to objects in Table \ref{tab:astromappend}. Light gray circles indicate no classified stars and most of them are field population (See Sect.~\ref{sec:infrared} for details). Particular objects described in Sect.~\ref{sec:p_objs} are identified. Green and blue curves are the MS (see text) shifted according to the adopted distance modulus with and without absorption/reddening, respectively. Red lines indicate the considered reddening path. Adopted reddening values for Region A/C and Region B are those presented in Table~\ref{tab:IRASComponents} for DBS~90-91 and DBS~89 respectively.}
\label{fig:phot}
\end{figure*}

Photometric color$-$color and color$-$magnitude diagrams (TCDs and CMDs) of the point objects located in Regions A and B are presented in  Fig.~\ref{fig:phot}a-f. All CMDs in each region clearly revealed the presence of differential reddening. We studied these diagrams and classified the objects following a procedure based on the computation of several reddening$-$free parameter values (see \citealt{bau20} for details), the photometric information at {\it Gaia} bands, and following the color conditions from WISE bands given by \cite{koe12}. Therefore, we could carry out a selection of different kinds of objects: a) early main sequence (MS) stars, b) objects with IR color excess and c) objects identified as class I and class II YSOs. The remaining objects were not classified and they were considered as field stars. 

In the above procedure we used, as a reference, the MS values given by \cite{sun13, koo83}, and the photometric relationships for {\it Gaia} filters indicated in its official Web page{\footnote{https://gea.esac.esa.int/archive/documentation/GEDR3/}}. We also considered a normal interstellar reddening law ($R_V = A_V / E_{B-V} = 3.1$) and the reddening model given by \cite{Wang-Chen2019}. The adopted extreme values for color excesses are presented in Table~\ref{tab:IRASComponents} as $E_{(B-V)min}$ and $E_{(B-V)max}$. Both border values were used to identify MS stars, while only the lower one was used to select objects with IR color excess (IR objects). Values for the other color indices ($E_{ICmin}$ and $E_{ICmax}$) were computed using the \cite{Wang-Chen2019} model. These border values are indicated also in the photometric diagrams by the location of the shifted MS ($E_{ICmin}$) and the point of the reddening vector ($E_{ICmax}$).

The upper MS has an almost vertical shape over the IR CMDs not allowing to estimate precise distance values. After trying several distances (from 2.2 to 3.4 kpc), we obtained the same objects with {\it Gaia} EDR3 photometric information as probable members for Region B. However, for each distance, we obtained a different set of objects with IR information as probable cluster members. Finally, we considered the distance value of 2.9 kpc obtained with the astrophotometric analysis (see Sect. \ref{sec:dist}). These CMDs provided also an estimation of the foreground color excess $E_{(B-V)min}$ of the studied young stellar populations. The excess values confirmed that Region\,A objects are much more reddened than Region\,B ones as was previously suggested by Fig.~\ref{fig:PhomIR}.

For these clusters, and following \cite{vig07} analysis, the spectral types of the adopted MS stars and IR objects were estimated de-reddening their positions in a $J$ vs $J-H$ diagram. Since this diagram avoids the use of $K$ band, possible IR excesses problems are minimized. However, this procedure depends on the precision with which the distance is known, providing in our case an approximate result. Therefore, we performed only a rought classification using the four labels O$^-$, O$^+$, B$^-$ and B$^+$, where O/B letters indicate O and B type stars, and -/+ signs mean early/late stars inside each type.

The amount of different kinds of objects identified for each studied cluster and the corresponding adopted spectral types are presented in Table \ref{tab:IRASComponents} and Table \ref{tab:pmandpRA}, respectively.

To look for YSO candidates in the surroundings of IRAS 14416$-$5937 source, we searched in a region centered at (l,b) = (316$^{\circ}$.8, -0$^{\circ}$.06) within a 0.1 degrees radius circle. We adopted the YSO candidates classification scheme described in \citet{koe12}, \citet{gut09} and \citet{lum02}, for the  WISE, GLIMPSE, and MSX data, respectively.

For the WISE sources, we used the WISE All$-$Sky Source Catalog \citep{Wri10}. We selected sources with photometric flux uncertainties lower than 0.2 mag and a signal$-$to$-$noise ratio greater than 7 in the W1, W2, and W3 bands. 
Following the criteria of \citet{koe12} for these sources, we first discarded the non$-$YSO sources with excess IR emission, such as PAH$-$emitting galaxies, broad$-$line active galactic nuclei (AGNs), unresolved knots of shock emission, and PAH$-$emission features. After that process, from the 248 sources from the list, only 34  were kept, of which we detected eight WISE Class I sources (i.e. sources where the IR emission arises mainly from a dense envelope, including flat spectrum objects). Regarding Class II sources (i.e. pre-main sequence stars with optically thick disks), we detected 26 candidates. As the last step, we also rejected one Class II source (J\,144552.74$-$59541) whose magnitudes satisfied ($W_1-W_3 <$ –1.7 $\times$ ($W_3-W_4$ + 4.3)), because according to \citet{koe12} it could be considered to not have sufficiently reddened colors, we obtained 25 class II sources.

The GLIMPSE sources were kept if their photometric uncertainties were lower than 0.2 mag. in all four IRAC bands (3.6, 4.5, 5.8, and  8.0 $\mu$m), taking these constraints into account, a total of 217 sources were selected. After applying the criteria of \cite{gut09} we detected four Class I  and ten Class II YSO candidates. All the detected YSO candidates but two were detected also by \cite{vig07}. It is worth mentioning that these authors carried out the search for YSOs in areas slightly different ($\sim$ 30\% bigger) from the one used in this work. \cite{sam18} detected the same four Class I YSO candidates.  
In addition, we  searched in the MSX Point Source Catalog (PSC; \citet{egan03}) looking for massive YSOs (MYSOs) and compact H\,II (CH\,II) regions candidates following the criteria of \cite{lum02}. We selected the sources with variability and reliability flags with values of 0 and flux qualities above 1 in all four MSX bands (8, 12, 14, and 21 $\mu$m). Only two sources from 32 fulfill these criteria. One of these is an MYSO (G316.8083$-$00.0500) and the other is a CH\,II candidate.
Neglecting the constraints of variability and reliability flags, we additionally identified the MSX MYSO candidate G316.8112$-$00.0566, already cataloged by \cite{bus06}. 
All WISE, GLIMPSE and MSX YSO candidates are listed in Tables \ref{YSOs1}, \ref{YSOs2}, and \ref{YSOs3}, respectively.

\begin{table*}
\begin{center}
\leavevmode
\caption{Main Parameters of the embedded clusters}
\label{tab:IRASComponents}
\begin{tabular}{lccccccc}
\hline
\hline
\multicolumn{1}{c}{ID} & $V_o - M_V$ & $E_{(B-V)min}$ & $E_{(B-V)max}$ & J$_{lim}$ & MS & IR & $U$ \\
& [mag] & [mag]& [mag] & [mag] & stars & objects &  [pc\,cm$^{-2}$] \\ 
\hline
DBS 89    & 12.24 & 2.5 & 5.5 & 16 & 11 & 6 & 116 \\
DBS 90-91 & 12.24 & 5.0 & 7.5 & 16 & 8 &  1 & 109 \\
\hline
\hline
\end{tabular}
\parbox{14cm}{}\\
\end{center}
\end{table*}

\begin{figure}
\centering
\includegraphics[trim=100 50 100 50, width=0.48\textwidth]{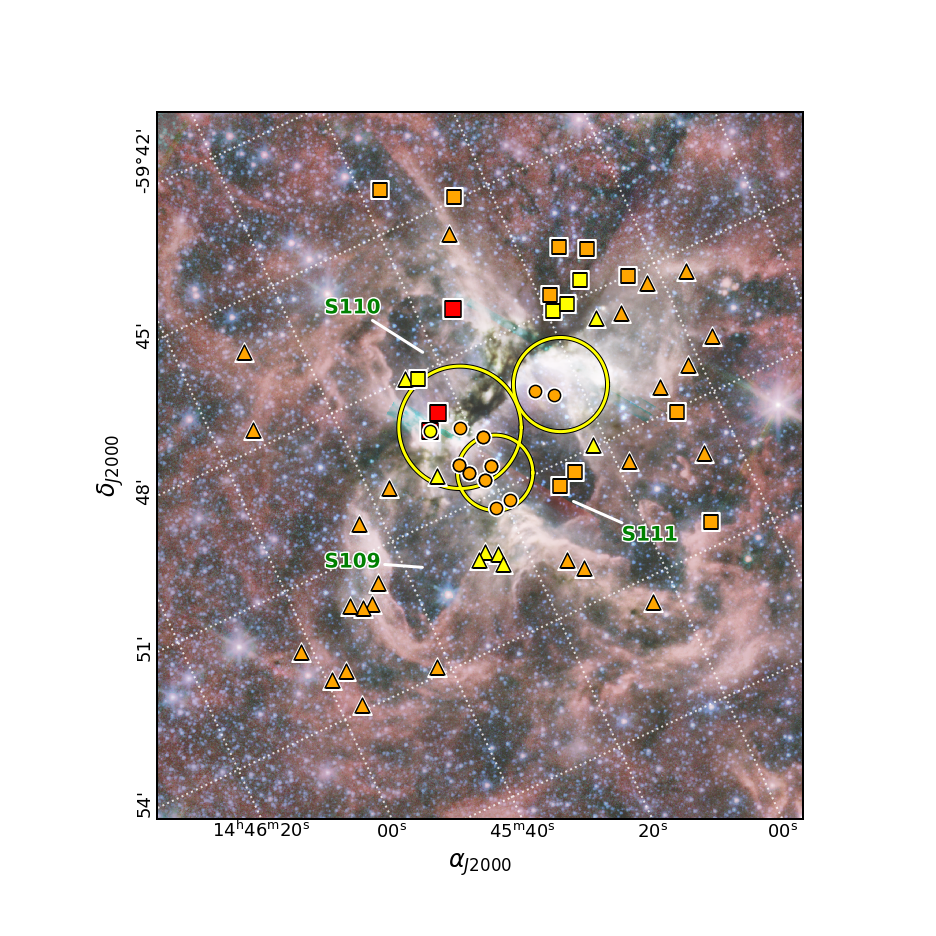}
\caption{Spitzer color image centered at G316.8$-$0.1 radio source (ch1 = violet, ch2 = cian, ch3 = yellow, ch4 = red). Large yellow hollow circles indicate Regions A, B and C presented in Fig.~\ref{fig:PhomIR}. Symbols are YSOs as presented in Fig.~\ref{fig:SUMSSAyB}. This is, yellow and orange symbols indicate Class I and II YSOs, respectively. Circles correspond to those identified using K band and WISE data, triangles those using only WISE data, and squares using GLIMPSE data. Big red squares indicate objects listed in Table \ref{YSOs3}. Bubbles S109, S110 and S111 presented by \cite{Churchwell2006} are identified.}
\label{fig:spitzer}
\end{figure}

\subsubsection{Astrometric analysis}
\label{sec:astrometry}

Astrometric data (position and proper motion) show that a star cluster is represented by an over$-$density in the sky as well as in the VPD. The astrometric method of analysis of these over$-$densities allows us to carry out an independent identification of their members and a determination of their astrometric parameters.

\begin{figure*}
\centering
\includegraphics[trim=100 10 100 10, width=0.7
\textwidth]{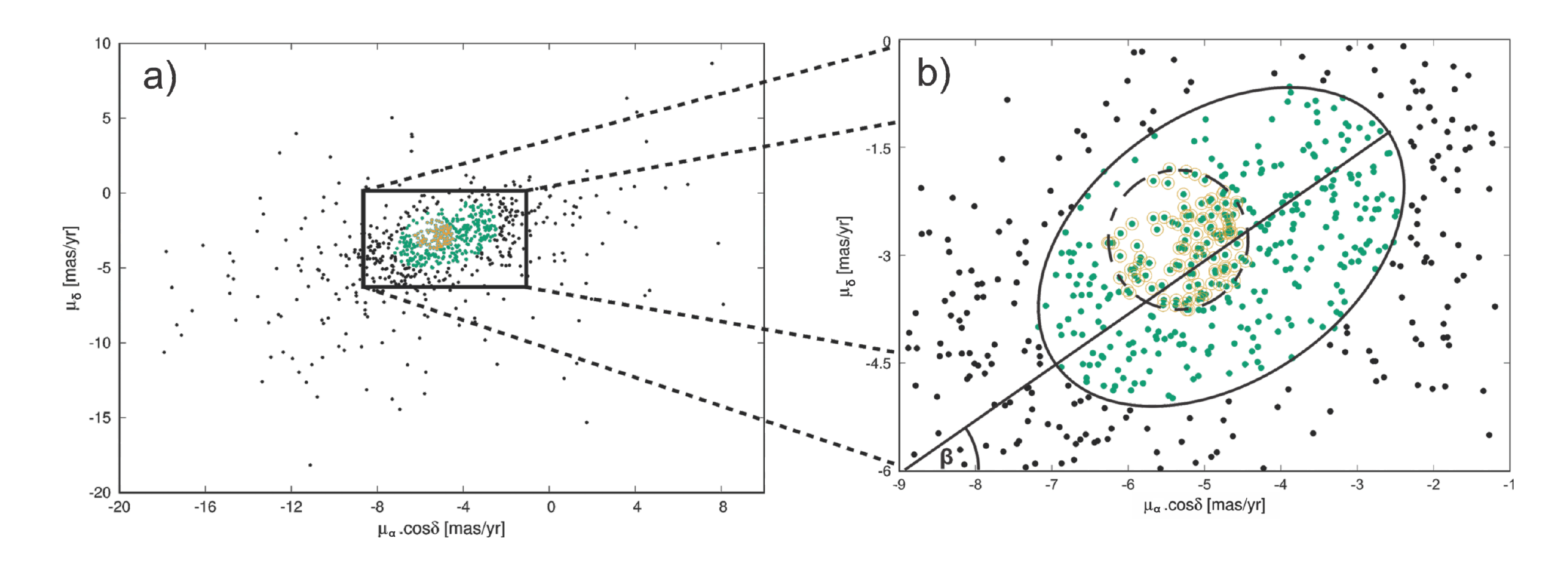}
\caption{
VPD of the 754 stars studied using astrometric $Gaia$ data. Panel b) present a zoomed view of the central part of panel a). Ellipse and slashed circle represent the enclosed curves for the selected objects for the field and clusters populations, respectively (see Sect \ref{sec:astrometry}). Green dots are the 353 stars selected from the {\it Gaia} EDR3 catalog and the yellow hollow symbols represent the 110 possible members obtained in this paper.}
\label{fig:PM}
\end{figure*}

First, we analyzed the VPD over$-$density adopting the mathematical model suggested by \citet{Vas58} and the technique based upon the maximum likelihood principle developed by \citet{San71}. This model assumes that the proper motion distribution $(\Phi_i(\mu_{xi},\mu_{yi}))$ of the region selected consists in the overlapping of two bivariate normal frequency functions.

\begin{eqnarray}
\label{phit}
\Phi_i(\mu_{xi},\mu_{yi}) = \phi_{1i}(\mu_{xi},\mu_{yi}) + \phi_{2i}(\mu_{xi},\mu_{yi}),
\end{eqnarray}

\noindent where $\phi_{1i}$ is a circular distribution for cluster stars, $\phi_{2i}$ is an elliptical distribution for field  stars, $\mu_{xi},\mu_{yi}$ are the i$-$th star 
proper motion in x and y, respectively. (see Fig. \ref{fig:PM}).

The circular and elliptical distributions take the following form

\begin{eqnarray}
\label{phi2}
&&\phi_{1i}(\mu_{xi},\mu_{yi})= \frac{N_c}{ 2\pi\sigma_{c}^2}\times \nonumber\\
&&\times\exp\left[-\frac{(\mu_{xi}-\mu_{xc})^2 + 
(\mu_{yi}- \mu_{yc})^2}{2\sigma_{c}^2}\right]
\end{eqnarray}

and
\begin{eqnarray}
\label{phi3}
&&\phi_{2i}(\mu_{xi},\mu_{yi})= \frac{N_f}{ 2\pi\sigma_{xf}\sigma_{yf}}\times\nonumber\\
&&\times\exp\left[-\frac{(\mu_{xi}-\mu_{xf})^2}{2(\sigma_{xf})^2} - 
\frac{(\mu_{yi}- \mu_{yf})^2}{2(\sigma_{yf})^2}\right],
\end{eqnarray}

\noindent where the symbols $\sigma_{xf}$, $\sigma_{yf}$ are the elliptical dispersion for the field stars, $\sigma_c$ the circular dispersion for the cluster stars, $\mu_{xf}$, $\mu_{yf}$ the field star mean proper motion, and $\mu_{xc}$, $\mu_{yc}$ the cluster mean proper motion. $N_c$ is the number of cluster members, and $N_f$ the number of field stars. These parameters were found by applying the maximum likelihood principle. Before applying the model, the VPD of the $N$ stars was rotated over angle $\beta$ to make its axes coincident with the field distribution ones (see Fig. \ref{fig:PM}.)

Then we could determine the membership probability for the i$-$th star from:

\begin{equation}
\label{Pc}
P_i(\mu_{xi},\mu_{yi})= \frac{\phi_{1i}(\mu_{xi},\mu_{yi})}{\Phi_i(\mu_{xi},\mu_{yi})}.
\end{equation}

Therefore, the $N_c$ stars with the highest $P_i(\mu_{xi},\mu_{yi})$ values are considered as members.

This method makes it possible to remove most of the field stars from the sample. The percentage of stars that have been removed depends on the distribution of the stars' proper motions. To know the degree of effectiveness we evaluated the membership using the index \textit{E} that follows \citet{sh96}.

\begin{equation}
\label{ef}
E= 1- \frac{N \sum_{i=1} ^ {N}  P_{i} [1- P_{i}] }{  \sum_{i=1} ^ {N} P_{i}   \sum_{i=1} ^ {N} [1-P_{i}]   }.
\end{equation}

When the moving group and the field stars are perfectly separated, we obtain the maximum value of {\it E} = 1, therefore \textit{E} is an estimate of the contamination of the selected sample. Therefore
we applied the analysis to the 754 stars selected from the {\it Gaia} EDR3 catalog. (see Sect. \ref{selection})
and we obtained the following parameters: $\mu_{\alpha}$cos${\delta}_c$, $\mu_{\delta_c}$, ${\sigma}_c$, $N_c$ (see Table~\ref{tabla_mu_pi_clas}). From equation \ref{Pc} we calculated the membership probability for each star and we identified the 110 probable proper motion members (see Fig.~\ref{fig:PM}).
The resolution of equation \eqref{ef} gave $E$ = 0.73 indicating a good separation between the field stars and those of the moving group.
 
\begin{table}
\caption{Astrometric parameters of the stellar groups}
\label{tabla_mu_pi_clas} 
\centering                                      
\begin{tabular}{lcccccccccr} 
\hline
\hline                         
\multicolumn{1}{c}{Stellar} & $\mu_{\alpha}$cos${\delta}_c$ & $\mu_{\delta_c}$ & ${\sigma}_c$ & $N_c$ \\ 
\multicolumn{1}{c}{group} & [mas\,yr$^{-1}$] & [mas\,yr$^{-1}$] & [mas\,yr$^{-1}$] & \\
\hline
$\phi_{1i}(\mu_{xi},\mu_{yi})$	 &  -5.18	$\pm$	0.06	& -2.87	$\pm$	0.06	& 0.48	$\pm$	0.01 &  110\\
 group B & 	-5.21	$\pm$ 	0.05	&	-2.81	$\pm$	0.05	&	 0.44	$\pm$	0.01 &  18\\
 group C & -5.23	$\pm$ 	0.06	&	-2.92	$\pm$	0.06	&	0.40	$\pm$	0.01 &  13\\
\hline                                      
\end{tabular}
\parbox{9cm}{Notes: Columns gives the mean proper motion components
($\mu_{\alpha}$cos${\delta}_c$, $\mu_{\delta_c}$),  the circular dispersion ($\sigma_c$) and number of members.}. 
\end{table}

In order to identify embedded clusters, we applied then the other astrometric condition that characterizes them, the over$-$density in the sky. 

\begin{figure}
\centering

\includegraphics[trim=50 40 70 50, width=0.43\textwidth]{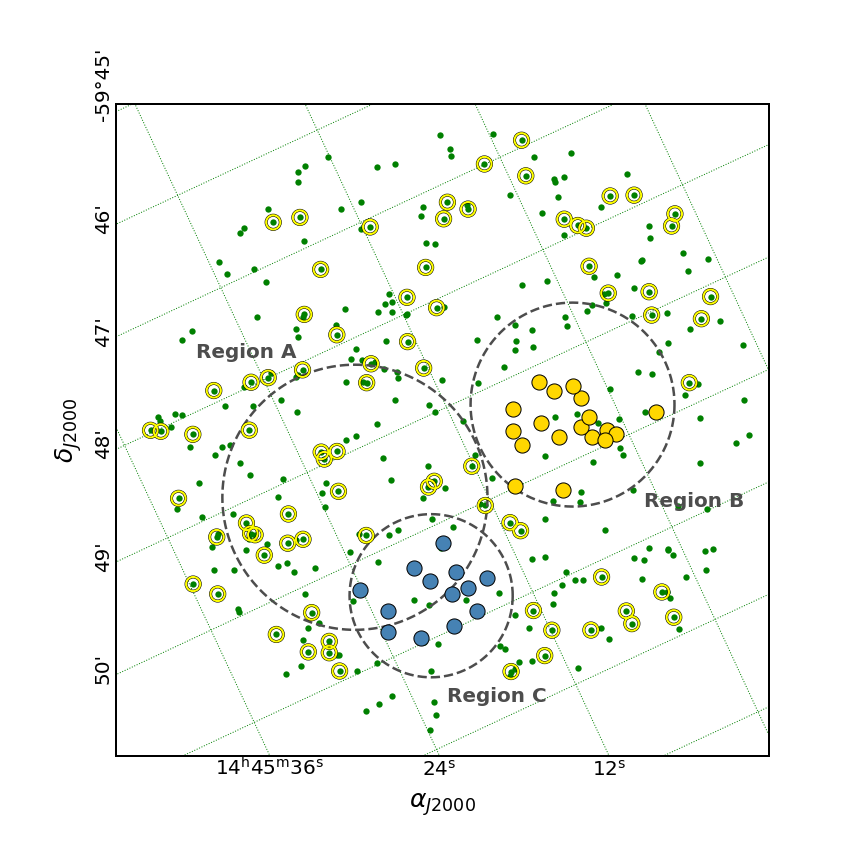}
\caption{Finding chart of probable astrometric members. Hollow yellow symbols and green dots have the same meaning that in Fig.~\ref{fig:PM}. Full yellow and blue symbols indicate the astrometric members of group~B, and group~C, respectively (see Sect. \ref{sec:astrometry}). Selected regions\,A, B, and  C (see Sect.\ref{sec:infrared}) are indicated by large slashed black circles.} 
\label{fig:Spitzer}
\end{figure}

In Fig. \ref{fig:Spitzer}, the position of stars following the gaussian elliptical distribution is presented with green filled symbols (353 stars)
and the 110 possible proper motion members are presented with yellow hollow symbols. The analysis of this chart allowed us to identify two possible stellar groups,
 field stars with the same proper motion. 
one of them denominated group\,B  with 18 stars (indicated with yellow filled symbols and  identified with the embedded cluster DBS\,89) and another denominated as group\,C  with 13 stars (represented with blue filled symbols). The astrometric members of each of these groups are presented in Table \ref{tab:astromappend} and their astrometric parameters in Table~\ref{tabla_mu_pi_clas}. The astrometric analysis did not detect any spatial over$-$density in Region\,A.

\subsubsection{DBS89 members}
\label{sec:DBS89members}

The comparison of the 18 possible astrometric members with the 19 possible photometric ones shows the following results for the DBS89 embedded cluster: a) there are nine stars in common between both analyzes that we have denominated astrophotometric members (see Table \ref{tab:parallax}), b) the photometric members  identified as 2MASS stars J14445838-5948416 and J14450022-5949489 in Table \ref{tab:pmandpRA} are not astrometric members, c) for the rest of the possible photometric members presented in Table \ref{tab:pmandpRA} there was not found any counterpart in {\it Gaia} EDR3 and d) the astrometric members  identified as "nm" in Table \ref {tab:astromappend} are not photometric members.

\subsubsection {Energetic balance}
\label{sec:energetic}

To study if the embedded clusters DBS\,89$-$90$-$91 could have generated the H\,II region, we computed the excitation parameter $\mu$ = $R_S$ $N_e^{2/3}$ of G316.8$-$0.1. This represents the amount of Lyman photons that had to be absorbed by the HI to originate the H\,II region (see Table~\ref{tab:AboutDistance}).

Additionally, the knowledge of the amount of Lyman photons emitted by the brightest stars of DBS\,89$-$90$-$91 is important to evaluate their capacity to generate the H\,II region. Therefore, we computed the ionization parameter ($U$) given by \citet{Wil13} as:
\begin{equation}
U = 3.241 \times 10^{-19}\,\sqrt[3]{\frac{3 \, N_{Ly}}
{ 4 \, \pi \, \alpha_2(T_e)}},
\end{equation}

\noindent where $\alpha_2(T_e)$ is the total recombination coefficient of hydrogen excluding captures to the ground level. The value of $\alpha_2(T_e)$ coefficient was computed using the expression of \citet{spi78} and the $T_e$ = 5400 K \citep{cas95}. For ionization parameter ($U$), the number of Lyman ionization photons provided by earlier stars than B$^+$ \citep{ster03} of possible members of DBS\,89$-$90$-$91 
was calculated. We considered atmosphere models for hot stars with solar metallicity and Luminosity Class V. We decided to use spectral type (ST) B\,1.5 for B$^-$ stars, ST = O\,8 for O$^+$ stars and ST = O\,5 for O$^-$ stars. Embedded cluster DBS\,89 (Region\,B) has eleven B$^-$, two O$^+$, and two O$^-$ stars, and all of them contribute a U value of $\sim$ 116 [pc cm$^{-2}$]. Embedded cluster DBS\,90$-$91 (Region\,A) has five B$^-$ stars and the particular Objs 1, 2, and 3 described in Sect.~\ref{sec:p_objs}. The spectral type of these Objs is possibly B$^-$ (Obj 1), O$^-$ (Obj 3), and Wolf$-$Rayet ($WR$) (Obj 2). \citet{law02} studied the ionizing fluxes of $WR$ stars and the Lyman continuum flux appears independent of the $WR$ spectral type being 47.44 $\leq$ Q $\leq$ 50.51. Since \citet{vig07} adopted for Obj 2 one ST earlier than O6 we decided to assign this the number of Lyman ionization photons provided by ST O\,5. In this way, DBS\,90$-$91 contribute to the interstellar medium with the Lyman ionization photons of six B$^-$ and two O$^-$ stars resulting a U value of $\sim$ 109 [pc cm$^{-2}$]. If we only considered the Lyman ionization photons of WR star as ST = O\,5, the U value would be $\sim$ 86 [pc cm$^{-2}$].  
After that, we compare parameters $\mu$ and U and we found U $>>$ $\mu$ being the relation $\frac{U}{\mu}$ = 1.2, if we take into account the WR star probable member of DBS\,90$-$91 and $\frac{U}{\mu}$ = 2.0, if we consider the DBS\,89 embedded cluster. \\

\section{Discussion}

\subsection{DBS89 cluster distance}
\label{sec:dist}

The distance of a star forming region is not easy to estimate, in particular in the fourth Galactic quadrant because of the ambiguity in distance.   
The research carried out in the G316.8$-$0.1 H\,II region, with observations obtained by different authors at different spectral ranges, proposed a distance value in the range from 2.5 \citep{cas87} to 3.3 kpc \citep{cas95}.
To resolve the kinematic distance ambiguity, \citet{bus06} searched some types of $^{13}$CO and H\,I spectral profiles of MYSOs observed with Mopra 22 millimetre wave telescope combined with SGPS H\,I data. 

Now we could estimate an astrometric distance to the DBS\,89 cluster using the \citet{bai21} distance catalog obtained from {\it Gaia} EDR3 parallaxes and selecting the 9 astrophotometric adopted members presented in Sect. \ref{sec:DBS89members}

Studies by \citet{bai21} in {\it Gaia} EDR3 show that for stars with a fractional error $f = \sigma_\pi/\pi$ $\leq$ 0.1, the inverse of parallax gives us a good estimate of the distance. For the cases in which it is greater, it is convenient to use the \citet{bai21} distance catalog where they use a three$-$dimensional model of the Galaxy, and include the interstellar extinction and the magnitude limit of {\it Gaia}. 
To infer the distance they develop two models: the geometric model where they use parallax and its error; and the photogeometric model where, in addition to parallax, they use the color and apparent magnitude of the star. We must take into account that the distances provided by \citet{bai21} in their catalog has incorporated the parallax zero point correction of $-$0.017 mas yr$^{-1}$ derived by \citet{lin21}.

According to \citet{bai21}, those distances corresponding to 0.1 $\leq f \leq$ 1 constitute the most important result of their work. 
However, the analysis shows that using data with a low fractional parallax error allows to restrict the uncertainty in distance.

Then, to select the best data we limit the $f$ value to $f \leq$ 0.5. Applying this condition we do not consider the first two stars in Table \ref{tab:parallax} for the distance calculation (EDR3 number 5878925503633099264  
and  5878925709791531776). 

Finally, to calculate the distance, we must choose between the two methods proposed by \citet{bai21} (geometric and photogeometric). Although we are in a region with high absorption, as all fractional errors are greater than 0.20, we considered that the incorporation of apparent magnitude and colour improves our final result.
Table \ref{tab:parallax} shows the median of the geometric distance posterior (rgeo) and the median of the photogeometric distance posterior (rpgeo). The value obtained for the average distance is rpgeo = 2.9 $\pm$ 0.5 kpc.
This value is in agreement with that obtained by \citet{bus06} resolving the ambiguity problem.

\subsection{Is group C associated with G316.8$-$0.1 H\,II region?}

Our astrometric analysis (see Sect.~\ref{sec:astrometry}) revealed a group of stars with similar proper motion values and lose spatial concentration. This group covers the indicated Region\,C which overlaps the already identified Region\,A. We calculated the distance to this group of stars taking into account the same selection criteria applied to determine the distance of DSB\,89 cluster. Then, of the 13 possible members, 6 were eliminated by $f \geq$ 0.5. The remaining 7 members provided an average distance value rpgeo = 3.1 $\pm$ 1.0 kpc. 
To evaluate the reliability of this group, we built the photometric diagrams of objects located in Region\,C (see Fig.~\ref{fig:phot}g-i). These diagrams show that all the astrometric identified stars in Region\,C 
are suffering a much lower reddening value than the adopted for the embedded clusters associated with G316.8$-$0.1 H\,II region. Therefore, this group C seems to belong to the foreground field population. 

\subsection{Particular objects}
\label{sec:p_objs}

\begin{figure}
\centering
\includegraphics[trim=100 30 100 50, width=0.48\textwidth]{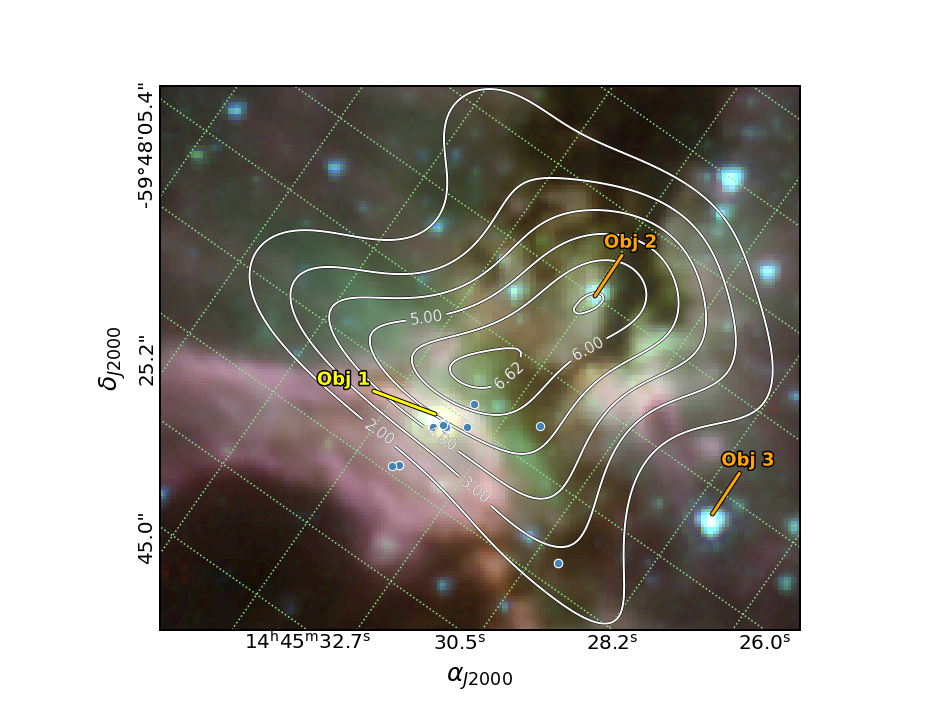}
\caption{Spitzer color image at the core of G316.8$-$0.1$-$A radio source (ch1 = blue, ch2 = green, ch3 = red) showing the three particular objects described in Sect.~\ref{sec:p_objs}. Light blue symbols indicate masers given by \cite{dal18}. White curves are radio continuum flux levels at 843 MHz (deconvolved SUMSS) revealing the presence of a double peak.}
\label{fig:spitzer_core}
\end{figure}

Some point objects deserve some special comments. They are those 2MASS identified as J14452143-5949251 (Obj 2), J14452450-5950084 (Obj 3), and J14452625-5949127 (Obj 1). The location of these objects is shown in detail in Fig.~\ref{fig:spitzer_core} and their main features are the following:
\begin{itemize}
\item Obj\,1: The spatial location of this object is only 14" eastward of the main peak of the radio source G316.8--0.1--A and is also almost coincident with the molecular clump C1 identified by \cite{sam18}. On the other hand, its location over the $JHK$ photometric diagrams is consistent with an early B$-$type star. It is not clear to identify the corresponding WISE and MSX counterparts for this object. We adopted the WISE J144526.30--594914.0 and MSX G316.8112--00.0566 (both at 1.3", see Table~\ref{YSOs3}) for this role. It must be noticed that in $K$ band images of VVV also appeared an elongated feature in this place. This feature was considered as a probable jet by \cite{sam18}. We classified the adopted WISE counterpart as a class I YSO and its MSX counterpart was considered as a massive YSO (MYSO) by \cite{bus06}. Additionally, the main peak of G316.8--0.1--A has an arc shape that resembles a bow$-$shock (see \citealt{MacLow1991}). Under this situation, this peak could be a consequence of UV radiation provided by the B$-$type star and the presence of an accretion disk \citep{Cesaroni2016}. This picture provides support to a monolithic star formation process given by \cite{Krumholz2009} since they found, using 3D simulations, that the radiation pressure of massive objects could not halt accretion. It must also be noticed that most of the masers in the H\,II region listed by \citet{dal18} are located in the surroundings of this object. Since maser emission is associated with a coherent velocity field as a disk, a jet, or an expanding shell \citep{Moscadelli2000}, this fact reinforces the previous scenario.
\item Obj 2: In this case, this object has a location coincident with the secondary peak of the radio source G316.8--0.1--A and the molecular clump C2 identified by \cite{sam18}. We classified its WISE counterpart (J144521.58--594925.3) as a class II YSO. This object has been identified as IRS10 by \cite{sha81} and considered as a probable O\,6 type star and the exiting source of all the H\,II region. A similar classification was achieved by \cite{vig07} that adopted is as a star earlier than O\,6. More recently, this object was considered by \cite{Anderson2014} as a WR candidate and a "colliding wind binary" (CWB). This conclusion was based on its location on the $JHK$ photometric diagrams and to be adopted as the counterpart of the ChI$\sim $J144519--5949${_2}$ source detected in the X$-$ray ChIcAGO survey.
\item Obj 3: This object has a similar location to the previous one over the $JHK$ photometric diagrams and its WISE counterpart (J144524.21--595007.6) was also classified as a class II YSO. It could be another WR star \citep{Mauerhan2011}, but in this location, there is neither a radio peak nor an X-ray emission. Therefore, we classified this object as a dubious early O-type star (O$^{-}$). 
\end{itemize}

\subsection{Are DBS\,89$-$90$-$91 the exciting clusters?}

In Sec \ref{sec:energetic} we presented the comparison between $U$ and $\mu$ parameters. The obtained $\mu$ parameter for G316.8$-$0.1$-$A radio source was $\mu$ = 71 pc\,cm$^{-2}$ and the number of Lyman ionization photons provided in the best case by WR star (Obj 2), probable member of DBS\,90$-$91, is $U$ = 86 pc\,cm$^{-2}$ 
resulting in this way $U > \mu$. This result shows that DBS\,90$-$91 embedded cluster with only the WR star could generate the G316.8$-$0.1$-$A H\,II region. 

G316.8$-$0.1$-$B H\,II region has $\mu$ = 57 pc\,cm$^{-2}$ and the number of Lyman ionization photons provided by earlier stars members of DBS\,89 is $U$ = 116 pc\,cm$^{-2}$, resulting in this way $U > \mu$.

Both results indicate that G316.8$-$0.1$-$A and G316.8$-$0.1$-$B are limited by density, and according to the expression (1 $-$ ($\frac{\mu}{U})^3$), there is an excess of $\sim$ 72\% of Lyman photons in the former case, and $\sim$ 88\% in the later. These photons are not ionizing the gas but they are absorbed by the dust and it is heated by them \citep{kur94}. 

With our investigation about the stellar members employing 2MASS, VVV and $Gaia$ EDR3 data we propose that the earliest stars of DBS\,89$-$90$-$91 clusters are responsible for the  G316.8$-$0.1 A and B regions.

The presence of YSOs  in  star-forming regions is indicative of ongoing star-formation activity. The study of filaments has attracted much attention being the original mater and host zone of star formation. Several studies have been carried out showing the role and importance that has the filaments in the star formation processes \citep[][]{sch14, zha19, zav20}. In our study, is striking the distribution of the YSOs shown in Figure \ref{fig:spitzer}. There can be appreciated that some of them are seen projected onto the  dark parental filament identified by \cite{sam18}, while most of them are seen projected onto the arcs of the 8 $\mu$m emission that comes from the photo-dissociated region (PDR). These findings are in agreement with that observed by other authors \citep[e.g.][]{bha22}.

\section{Conclusions}

In this work, we carried out a multiwavelength study of the DBS\, 89--90--91 embedded clusters region, 
we determined their stellar members and their interaction with the ISM.
From an astrophotometric analysis, we have confirmed that DBS\, 89, linked to G316.8--0.1 radio source, 
is located at a distance of 2.9 $\pm$ 0.5 kpc from the Sun.

From the photometric analysis of the region, we identified 26  OB-type star member candidates, ten Class II, and one Class I YSOc. Among the 26 stars, nine would be astrophotometric members of DBS\, 89, eight  would be only photometric members of DBS\, 89, and the remaining nine sources would be photometric members of DBS\, 90--91.

We detected several Class I and Class II YSOcs, two MYSOs, and one CH~II distributed throughout the G316.8--0.1 radio source. Some of them are seen preferably projected onto the photodissociated region and others are seen projected onto the dark parental filament. The presence of several YSOs located in the regions under study allows us to conclude this is an active star-forming region.

We have also estimated the main physical parameters of the H\,II region and identified the earliest-type stars of the embedded clusters as their main exciting sources.  With all these results was possible to improve the knowledge about the stellar components present in the Sagittarius-Carina arm of our Galaxy and its interaction with the ISM.

\begin{acknowledgements}
We wish to thank our referee's suggestions and comments, which improved   the original version of this work. This research has received financial support from UNLP "Programa de Incentivos" 11/G158, 11/G168 and 11/G172, and UNLP PPID\,G005; CONICET PIPs\,112-201701-00507CO, 112-201701-00055 and 112-201701-00604, and Agencia I+D+i PICT 2019-0344. The authors want to thank for the use of the NASA Astrophysics Data System, of the SIMBAD database and ALADIN tools (CDS, France). This publication was based on: a) Observations and data products from observations made with ESO Telescopes at La Silla or Paranal Observatories; b) The 2MASS, which is a joint project of the University of Massachusetts and the IPAC/California Institute of Technology, funded by the NASA and the NSF; c) Data products from the WISE, which is a joint project of the University of California, Los Angeles, and the JPL/California Institute of Technology, and is funded by the NASA; d) Observations made with the Spitzer Space Telescope, which is operated by the Jet Propulsion Laboratory, California Institute of Technology under a contract with NASA; e) Data products from the Midcourse Space Experiment. Processing of the data was funded by the Ballistic Missile Defense Organization with additional support from NASA Office of Space Science. This research has also made use of the NASA/ IPAC Infrared Science Archive, which is operated by the Jet Propulsion Laboratory, California Institute of Technology, under contract with the National Aeronautics and Space Administration; f) The SGPS is a project with images obtained at high resolution using the Australia Telescope Compact Array and the Parkes Radio Telescope; g) The SUMSS is a radio imaging survey carried out with the Molonglo Observatory Synthesis Telescope (MOST), Australia and g) This work has made use of data from the European Space Agency (ESA) mission {\it Gaia} (\url{https://www.cosmos.esa.int/gaia}), processed by the {\it Gaia} Data Processing and Analysis Consortium (DPAC,\url{https://www.cosmos.esa.int/web/gaia/dpac/consortium}). Funding for the DPAC has been provided by national institutions, in particular the institutions participating in the {\it Gaia} Multilateral Agreement.
\end{acknowledgements}

\bibliographystyle{aa}  
\bibliography{biblioAA2019}

\begin{appendix}

\section {Photometric members}

\begin{table*}
\begin{center}
\leavevmode
\begin{scriptsize}
\caption{Probable members of embedded clusters DBS\,89 (Region\,B) and DBS\,90$-$91 (Region\,A$+$C) from photometric observations. Equatorial coordinates, $JHK$ magnitudes, approximate spectral classification and adopted object types are presented.}
\label{tab:pmandpRA}
\begin{tabular}{lccccccccc}

\hline
\hline
Cluster & ID & $\alpha_{J2000}$ & $\delta_{J2000}$ & J & H & K & SpT & Obj.~type & {\bf Notes}\\
 & & [h:m:s] & [$\circ$:':''] &  [mag] & [mag] & [mag] & & & \\
\hline
DBS\,89
 &  2MASS J14445838-5948416 & 14:44:58.4 & -59:48:41.7 & 15.06 & 13.85 & 12.93 & B$^+$ &     IR & Gaia          \\
 &  2MASS J14450022-5949489 & 14:45:00.2 & -59:49:49.0 & 14.30 & 12.90 & 12.31 & B$^-$ &     MS & Gaia          \\
 &  2MASS J14450056-5949055 & 14:45:00.6 & -59:49:05.7 & 13.17 & 12.40 & 12.05 & B$^-$ &     MS & Gaia \\
 &  2MASS J14450217-5949281 & 14:45:02.2 & -59:49:28.3 & 14.21 & 13.17 & 12.64 & B$^-$ &     MS & Gaia$^{\ast}$ \\
 &  2MASS J14450222-5949405 & 14:45:02.2 & -59:49:40.7 & 12.35 & 10.91 & 10.36 & B$^-$ &     MS & Gaia$^{\ast}$ \\
 &  2MASS J14450240-5949199 & 14:45:02.4 & -59:49:20.2 & 13.39 & 12.48 & 12.03 & B$^-$ &     MS & Gaia$^{\ast}$ \\ 
 &  2MASS J14450275-5949302 & 14:45:02.7 & -59:49:30.3 & 15.21 & 12.93 & 10.59 & O$^-$ &     IR &               \\
 &  2MASS J14450306-5949518 & 14:45:03.1 & -59:49:52.0 & 15.67 & 14.06 & 13.09 & B$^+$ &     IR &               \\
 &  2MASS J14450320-5949440 & 14:45:03.2 & -59:49:44.2 & 11.94 & 10.86 & 10.34 & B$^-$ &     MS & Gaia$^{\ast}$ \\
 &  2MASS J14450460-5949095 & 14:45:04.6 & -59:49:09.9 & 13.73 & 12.66 & 12.18 & B$^-$ &     MS & Gaia$^{\ast}$ \\
 &  2MASS J14450455-5949373 & 14:45:04.6 & -59:49:38.4 & 13.35 & 11.92 & 10.37 &    -- & YSO II &               \\
 &  2MASS J14450497-5949286 & 14:45:05.0 & -59:49:28.4 & 15.51 & 13.09 & 11.41 & B$^-$ &     IR &               \\ 
 &  2MASS J14450498-5949434 & 14:45:05.0 & -59:49:44.0 & 11.78 & 10.17 & ~9.49 & O$^+$ &     MS & Gaia$^{\ast}$ \\
 &  2MASS J14450518-5948498 & 14:45:05.2 & -59:48:49.9 & 15.71 & 13.54 & 10.95 & O$^-$ &     IR &               \\
 &  2MASS J14450519-5949046 & 14:45:05.2 & -59:49:04.5 & 14.85 & 13.12 & 11.94 & B$^-$ &     IR &               \\
 &  2MASS J14450584-5949317 & 14:45:05.8 & -59:49:31.9 & 11.83 & 10.22 & ~9.44 & O$^+$ &     MS & Gaia$^{\ast}$ \\
 &  2MASS J14450647-5950131 & 14:45:06.5 & -59:50:13.3 & 13.96 & 12.99 & 12.57 & B$^-$ &     MS & Gaia$^{\ast}$ \\
 &  2MASS J14450714-5949243 & 14:45:07.3 & -59:49:23.7 & 16.97 & 14.99 & 13.80 &    -- & YSO II &               \\
 &  2MASS J14450973-5949588 & 14:45:09.7 & -59:49:59.0 & 13.46 & 12.50 & 12.12 & B$^-$ &     MS & Gaia$^{\ast}$ \\
\hline 
DBS\,90 -- 91
 &  2MASS J14451515-5950030 & 14:45:15.2 & -59:50:02.9 & 16.87 & 14.69 & 13.69 & B$^+$ &     MS & \\
 &  2MASS J14451625-5950578 & 14:45:16.3 & -59:50:57.6 & 16.49 & 14.25 & 13.14 & B$^-$ &     MS & \\
 &  2MASS J14451843-5949499 & 14:45:18.6 & -59:49:49.2 & 15.46 & 12.64 & 10.62 &    -- & YSO II & \\
 &  2MASS J14451858-5949434 & 14:45:18.6 & -59:49:43.4 & 15.13 & 13.27 & 12.31 & B$^-$ &     MS & \\
 &  2MASS J14451892-5950201 & 14:45:18.9 & -59:50:20.1 & 16.13 & 14.39 & 13.52 & B$^+$ &     MS & \\
 &  2MASS J14451905-5951161 & 14:45:19.0 & -59:51:15.9 & 16.27 & 12.57 & 10.53 &    -- & YSO II & \\
 & WISE J144519.27-595026.8 & 14:45:19.4 & -59:50:27.3 & 18.58 & 16.54 & 15.33 &    -- & YSO II & \\ 
 &  2MASS J14451967-5951002 & 14:45:19.7 & -59:51:00.1 & 17.08 & 14.68 & 13.57 & B$^+$ &     MS & \\
 &  2MASS J14452039-5950210 & 14:45:20.4 & -59:50:21.0 & 16.99 & 14.74 & 13.52 & B$^+$ &     MS & \\
 &  2MASS J14452137-5950387 & 14:45:21.4 & -59:50:39.0 & 17.59 & 13.84 & 11.75 &    -- & YSO II & \\
 &  2MASS J14452143-5949251 & 14:45:21.5 & -59:49:25.3 & 14.12 & 10.95 & ~9.30 & WR - CWB (?) & YSO II & Obj2 / C2 \\
 &  2MASS J14452167-5951185 & 14:45:21.7 & -59:51:18.5 & 16.49 & 13.62 & 11.56 &    -- & YSO II & \\
 &  2MASS J14452265-5949521 & 14:45:22.7 & -59:49:52.2 & 15.85 & 14.07 & 12.70 & B$^-$ &     IR & \\
 &  2MASS J14452336-5950224 & 14:45:23.4 & -59:50:23.1 & 15.83 & 13.14 & 11.69 &    -- & YSO II & \\
 &  2MASS J14452373-5949375 & 14:45:23.7 & -59:49:37.5 & 15.46 & 13.57 & 12.61 & B$^-$ &     MS & \\
 &  2MASS J14452450-5950084 & 14:45:24.4 & -59:50:08.0 & 14.26 & 10.43 & ~8.58 & O$^- (?)$ & YSO II & Obj3 \\ 
 &  ~~2MASS J14452625-5949127$^{**}$ & 14:45:26.3 & -59:49:12.8 & 14.84 & 12.72 & 11.60 & B$^- (?)$ & YSO I & Obj1 / C1 \\
 &  2MASS J14452664-5949118 & 14:45:26.6 & -59:49:11.9 & 14.68 & 12.80 & 12.02 & B$^-$ &     MS & \\
\hline
\end{tabular}
\parbox{16cm}{Notes: 
"SpT": Approximate spectral classification from photometric data (see Sect.~\ref{sec:infrared} for details). \\
"Obj. type": Adopted object nature from our photometric analysis (see Sect.~\ref{sec:p_objs} for details). \\
"Notes": "Gaia" indicates objects with GAIA EDR3 information. In particular, $^{\ast}$ symbol indicates those probable astrometric cluster members detailed in Table~\ref{tab:parallax}. Obj1, Obj2 and Obj3 indicate the particular objects described in Sect.~\ref{sec:p_objs} and showed in Fig.~\ref{fig:spitzer_core}.  C1/C2 indicate objects located at radio source G316.8$-$0.1$-$A peak and probably associated, respectively, with C1/C2 molecular clumnps identified by \cite{sam18}.
} \end{scriptsize}
\end{center}
\end{table*}

\section{Young Stellar Objects}

\begin{table*}
\leavevmode
\begin{center}
\caption {YSO candidates identified from WISE catalog.}
\label{YSOs1}
\begin{tabular}{lccrrrrl}
\hline
\hline
ID & \multicolumn{1}{c}{$\alpha_{J2000}$} & \multicolumn{1}{c}{$\delta_{J2000}$} & 
              \multicolumn{1}{c}{$W_1$} & \multicolumn{1}{c}{$W_2$} & \multicolumn{1}{c}{$W_3$} & \multicolumn{1}{c}{$W_4$} & Class\\
            & \multicolumn{1}{c}{[h:m:s]} & \multicolumn{1}{c}{[$^\circ$:$\arcmin$:"]} & 
              \multicolumn{1}{c}{[mag]} & \multicolumn{1}{c}{[mag]} & \multicolumn{1}{c}{[mag]} & \multicolumn{1}{c}{[mag]} & \\
\hline

J144452.89-594832.2	&	14:44:52.9	&	-59:48:32.3	&	9.34	&	8.33	&	2.00	&	-1.59	& I  	\\
J144502.32-595057.9	&	14:45:02.3 &	-59:50:58.0	&	10.81	&	9.67	&	2.02	&	-2.16	& I	\\
J144524.67-595226.2	&	14:45:24.7	&	-59:52:26.2 &	11.63	&	10.52	&	4.93	&	0.84	& I	\\
J144524.71-595212.5	&	14:45:24.7 &	-59:52:12.6	&	11.37	&	10.36	&	4.68	&	2.04	&  I	\\
J144526.42-594759.4	&	14:45:26.4	&	-59:47:59.5	&	11.77	&	10.62	&	2.46	&	1.85	& I	\\
J144526.54-595202.4	&	14:45:26.5	&	-59:52:02.5	&	11.65	&	10.51	&	5.91	&	0.85	& I	\\
J144528.02-595208.8	&	14:45:28.0	&	-59:52:08.8	&	11.37	&	10.27	&	5.05	&	0.86	& I	\\
J144528.43-595009.1	&	14:45:28.4	&	-59:50:09.1	&	6.31	&	5.29	&	-0.52	&	-3.59	& I	\\

J144435.72-594826.3	&	14:44:35.7 &	-59:48:26.4	&	10.92	&	10.50	&	5.86	&	4.14	& II	\\
J144436.33-594954.8	&	14:44:36.3	&	-59:49:54.8	&	11.58	&	10.66	&	6.61	&	4.33	& II	\\
J144442.06-595016.5	&	14:44:42.1	&	-59:50:16.6	&	9.43	&	8.48	&	6.94	&	2.20	& II	\\
J144442.44-594819.2	&	14:44:42.4	&	-59:48:19.2	&	11.06	&	10.57	&	5.96	&	4.05	& II	\\
J144445.82-595206.6	&	14:44:45.8	&	-59:52:06.6	&	10.25	&	9.94	&	6.67	&	2.03	& II	\\
J144447.96-595026.2	&	14:44:48.0	&	-59:50:26.2	&	9.72	&	9.31	&	4.53	&	0.62	& II	\\
J144448.57-594840.4	&	14:44:48.6	&	-59:48:40.4	&	10.06	&	9.16	&	4.79	&	0.86	&	 II \\
J144458.00-595134.9	&	14:44:58.0	&	-59:51:35.0	&	10.59	&	10.23	&	5.51	&	0.76	&	 II \\
J144504.33-595431.3	&	14:45:04.3	&	-59:54:31.4	&	11.07	&	10.55	&	5.81	&	2.62	&	II \\
J144509.35-594535.5	&	14:45:09.3	&	-59:45:35.5	&	10.76	&	10.23	&	5.34	&	3.18	&	II \\
J144512.42-595314.6	&	14:45:12.4	&	-59:53:14.6	&	10.60	&	10.14	&	5.39	&	0.88	&	II \\
J144514.56-595256.6	&	14:45:14.6	&	-59:52:56.6	&	8.88	&	8.24	&	5.16	&	1.08	&	II \\
J144536.69-594956.4	&	14:45:36.7	&	-59:49:56.4	&	7.08	&	6.32	&	4.96	&	2.01	&	II \\
J144542.16-595349.0	&	14:45:42.2	&	-59:53:49.1	&	9.00	&	8.56	&	7.31	&	2.67	&	II \\
J144543.72-595021.8	&	14:45:43.7	&	-59:50:21.9	&	10.21	&	9.82	&	5.01	&	1.10	&	II \\
J144545.22-595140.9	&	14:45:45.2	&	-59:51:40.9	&	11.64	&	10.98	&	6.11	&	3.47	&	II \\
J144547.61-595201.1	&	14:45:47.6	&	-59:52:01.2	&	10.25	&	9.64	&	5.16	&	2.57	&	II \\
J144549.11-594601.7	&	14:45:49.1	&	-59:46:01.8	&	8.13	&	7.78	&	5.77	&	3.62	&	II \\
J144549.31-595201.2	&	14:45:49.3	&	-59:52:01.2	&	11.07	&	10.61	&	5.98	&	2.17	&	II \\
J144551.08-595151.8	&	14:45:51.1	&	-59:51:51.9	&	12.34	&	11.82	&	7.32	&	3.67	&	II \\
J144553.25-594736.1	&	14:45:53.2	&	-59:47:36.1	&	9.33	&	8.95	&	6.73	&	2.87	&	II \\
J144556.29-595304.6	&	14:45:56.3	&	-59:53:04.7	&	11.26	&	10.91	&	7.94	&	4.75	&	II \\
J144556.32-595352.8	&	14:45:56.3	&	-59:53:52.8	&	11.77	&	11.15	&	6.96	&	5.54	&	II \\
J144559.05-595307.1	&	14:45:59.0	&	-59:53:07.2	&	11.10	&	10.68	&	6.66	&	4.60	&	II \\
J144601.88-595218.8	&	14:46:01.9	&	-59:52:18.8	&	11.16	&	10.74	&	8.22	&	5.95	&	II \\

\hline
\end{tabular}
\end{center}
\end{table*}

\begin{table*}
\leavevmode
\begin{center}
\caption {YSO candidates identified from GLIMPSE catalog.}
\label{YSOs2}
\begin{tabular}{l l l r r r r l}
\hline
\hline
ID & \multicolumn{1}{c}{$\alpha_{J2000}$} & \multicolumn{1}{c}{$\delta_{J2000}$} & 
                 \multicolumn{1}{c}{3.6 $\mu$m} & \multicolumn{1}{c}{4.5 $\mu$m} & \multicolumn{1}{c}{5.8 $\mu$m} & \multicolumn{1}{c}{8 $\mu$m} & Class\\
SSTGLMC...     & \multicolumn{1}{c}{[h:m:s]} & \multicolumn{1}{c}{[$^\circ$:$\arcmin$:"]} & 
                 \multicolumn{1}{c}{[mag]} & \multicolumn{1}{c}{[mag]} & \multicolumn{1}{c}{[mag]} & \multicolumn{1}{c}{[mag]} & \\
\hline
G316.7582-00.0030*+	&	14:44:52.6 &	-59:47:39.4	&	14.62	&	12.51	&	11.13	&	10.24	& I	 \\
G316.7627-00.0115*+	&	14:44:56.3 &	-59:48:00.3 &	10.95	&	8.88	&	7.72	&	6.68	& I	 \\
G316.7676-00.0141*+	&	14:44:58.9 &	-59:48:01.3 &	14.28	&	11.86	&	10.69	&	10.34	& I	 \\
	G316.8152-00.0380*+	& 14:45:24.3 & -59:48:06.7 &	12.02	&	10.94	&	9.88	&	9.06	& I	 \\
G316.7409-00.0014*	&	14:44:44.9	&	-59:48:00.5	&	11.62	&	11.10	&	10.67	&	10.25	& II	 \\	
G316.7235-00.0497*	&	14:44:47.1 & -59:51:04.4 &	9.50	&	9.24	&	8.69	&	8.36	&	 II \\
G316.7557+00.0079*	&	14:44:49.3 & -59:47:07.5	&	12.98	&	12.02	&	11.34	&	10.79	&	 II \\
G316.7113-00.0885	&	14:44:49.6 & -59:53:29.6	&	11.63	&	11.08	&	10.35	&	9.31	&	 II	 \\
G316.7655+00.0087*	&	14:44:53.4 &	-59:46:49.8 &	10.32	&	9.84	&	9.36	&	8.80	&	 II	 \\
G316.7687-00.0084*	&	14:44:58.3 &	-59:47:41.2 &	12.88	&	12.09	&	11.58	&	10.67	& II	 \\
G316.8026+00.0261*	&	14:45:05.9 &	-59:44:56.4	&	11.97	&	11.72	&	11.28	&	9.82	& II	 \\
G316.7595-00.0710*	&	14:45:07.0 &	-59:51:19.3	&	12.12	&	11.53	&	11.16	&	10.65	& II	 \\
G316.7651-00.0760*	&	14:45:10.4 &	-59:51:27.1	&	11.09	&	10.57	&	10.08	&	9.60	& II	 \\
G316.8288+00.0288	&	14:45:16.7 &	-59:44:07.6	&	11.74	&	11.08	&	10.30	&	9.56	& II	 \\
\hline
\end{tabular}
\parbox{14cm}{{Notes: * detected by \cite{vig07}} \\
{+detected by \cite{sam18}}}\\
\end{center}
\end{table*}

\begin{table*}
\leavevmode
\begin{center}
\caption {YSO candidates and one CH\,II region identified from MSX catalog sources.}
\label{YSOs3}
\begin{tabular}{lccrrrr l}
\hline
\hline

\multicolumn{1}{c}ID & $\alpha_{J2000}$ & $\delta_{J2000}$ & 
             \multicolumn{1}{c}{8 $\mu$m}  & \multicolumn{1}{c}{12 $\mu$m} & \multicolumn{1}{c}{14 $\mu$m} & \multicolumn{1}{c}{21 $\mu$m} &  Type\\
           & [h:m:s] & [$^\circ$:$\arcmin$:"] & 
             \multicolumn{1}{c}{[Jy]} & \multicolumn{1}{c}{[Jy]} & \multicolumn{1}{c}{[Jy]} & \multicolumn{1}{c}{[Jy]} & \\
\hline 
 G316.8083-00.0500** &	14:45:23.8 &	-59:48:56.5 &	3.93 &	28.79 &	48.24	& 106.27 & MYSO\\
 G316.8112-00.0566++ &	14:45:26.4 &	-59:49:13.4 &	14.05 &	35.74 &	41.55 &	188.47 & MYSO\\
 G316.8028-00.0131 &	14:45:14.0 &	-59:47:04.9 &	0.87 &	2.11 &	1.48 &  4.87 & CHII\\

\hline
\end{tabular}
\parbox{14cm}{Notes:
{** detected in PSC singleton catalogs. Consist of sources that were detected once when multiple coverages of a field were taken.}\\
{++ Detected in the PSC when not considering  conditions of the variability and reliability flags. Identified by \cite{bus06}}}\\
\end{center}
\end{table*}

\section{Individual stellar proper motions}

\longtab[1]{
\begin{longtable}{ccccccccc}
\caption{Adopted astrometric members for group B and group C.}
\label{tab:astromappend} \\
\hline
\hline
ID GAIA EDR3 & $\alpha_{J2000}$ & $\delta_{J2000}$ & $\mu_{\alpha}\,cos \delta$ & $\mu_{\delta}$ & $G$ &  Group & Probability & Photometric\\
 58789................ &  [h:m:s] & [$\circ$:':'']  & [mas\,yr$^{-1}$] & [mas\,yr$^{-1}$] & [mag] &  &  & member\\
\hline
\endfirsthead
\caption{(continuation)}\\
\hline 
ID GAIA EDR3 & $\alpha$ & $\delta$ & $\mu_{\alpha}\,cos \delta$ & $\mu_{\delta}$ & G  &  group & Probability \\
 58789................ &  [h:m:s] & [$\circ$:':'']  & [mas\,yr$^{-1}$] & [mas\,yr$^{-1}$] & [mag] & & \\
 \hline
\endhead
 
24777773276800 & 14:45:18.6 & -59:50:58.4 & -4.647 & -2.677 & 18.7 & C & 0.88 & nm\\
24713359096320 & 14:45:21.4 & -59:50:56.6 & -5.691 & -3.461 & 17.2 & C & 0.80 & nm\\
24777773273344 & 14:45:16.5 & -59:50:56.3 & -5.333	& -3.144 & 18.7 & C  & 0.92 & nm\\
24747718835584 & 14:45:23.6 & -59:50:45.0 & -5.043 & -2.51 & 15.8 & C & 0.92 & nm\\
24782066273280 & 14:45:16.4 & -59:50:41.8 & -5.177 & -2.446 & 20.0 & C & 0.91 & nm\\
24679003008768 & 14:45:14.8 & -59:50:41.1 & -4.911 & -2.450 & 19.5 & C & 0.90 & nm\\
24782067303680 & 14:45:17.7 & -59:50:40.8 & -5.837 & -3.030 & 20.0 & C & 0.85 & nm\\
24812133023488 & 14:45:22.9 & -59:50:34.1 & -5.968 & -3.199 & 17.6 & C & 0.75 & nm\\
24782066278016 & 14:45:16.8 & -59:50:30.4 & -5.421 & -2.541 & 20.2 & C  & 0.91 & nm\\
25499327763456 & 14:45:06.5 & -59:50:13.3 & -4.661	& -2.698 & 18.6 & B & 0.89 & m\\
24777773277184 & 14:45:18.9 & -59:50:28.4 & -4.810 & -2.715 & 20.2 & C & 0.91 & nm\\
24777770939008 & 14:45:19.6 & -59:50:17.4 & -5.294 & -2.888 & 19.0 & C & 0.94 & nm\\
24816438317952 & 14:45:24.1 & -59:50:15.7 & -4.707 & -2.455 & 18.8 & C & 0.86 & nm\\
25537992826496 & 14:45:16.7 & -59:50:11.3 & -5.082 & -3.614 & 14.6 & C & 0.81 & nm\\
25503620808704 & 14:45:09.7 & -59:49:59.0 & -5.201 & -3.611 & 18.2  & B & 0.82 & m\\
25671126446592 & 14:45:01.9 & -59:49:57.0 & -5.478 & -1.997 & 20.3  & B & 0.70 & nm\\
25671126444544 & 14:45:00.9 & -59:49:56.6 & -5.526	& -2.472 & 18.4  & B & 0.81 & nm\\
25671122242176 & 14:44:02.7 & -59:49:54.6 & -5.623 & -2.642 & 17.8  & B & 0.72 & nm\\
25671124141696 & 14:45:01.4 & -59:49:52.4 & -4.714 & -3.433 & 19.0  & B & 0.93 & nm\\
25499323551104 & 14:45:02.7 & -59:49:51.9 & -5.032 & -3.604 & 17.8  & B & 0.91 & nm\\
25709791805824 & 14:45:03.2 & -59:49:44.2 & -4.927 & -2.627 & 16.9  & B & 0.73 & m\\
25503620821120 & 14:45:05.0 & -59:49:44.0 & -4.759 & -2.148 & 18.2  & B & 0.69 & m\\
25709779257984 & 14:45:02.2 & -59:49:40.7 & -5.220 & -2.454 & 17.5  & B & 0.82 & m\\
25503622488832 & 14:45:07.9 & -59:49:39.0 & -5.950 & -2.562 & 20.7  & B & 0.89 & nm\\
25503633099264 & 14:45:05.8 & -59:49:31.9 & -4.745 & -2.115 & 18.5  & B & 0.90 & m\\
25499323563648 & 14:45:08.0 & -59:49:29.3 & -5.917 & -3.282 & 18.1  & B & 0.88 & nm\\
25709791531776 & 14:45:02.2 & -59:49:28.3 & -4.799 & -2.714 & 19.6  & B & 0.79 & m\\
25709791532928 & 14:45:02.4 & -59:49:20.1 & -4.675 & -2.626 & 17.9  & B & 0.73 & m\\
25705486188544 & 14:45:03.8 & -59:49:18.5 & -5.348 & -2.571 & 19.4  & B & 0.78 & nm\\
25499323569152 & 14:45:07.4 & -59:49:17.7 & -5.258 & -2.793 & 17.0  & B & 0.83 & nm\\
25705486190208 & 14:45:04.6 & -59:49:09.9 & -4.909 & -2.216 & 19.0  & B & 0.89 & m\\
\hline
 \end{longtable}
 \parbox{18cm} {Notes: Columns gives  the {\it Gaia} EDR3 identifier, position $(\alpha,\delta)_{J2000}$, proper motion components $(\mu_\alpha cos\delta, \mu_\delta)$ and $G$ magnitude from {\it Gaia} EDR3, belonging region, probability and photometric members.}\\
 }
 
\begin{table*}
\begin{center}
\caption{{\it Gaia} EDR3 astrometric parameters of  DBS\,89 astrophotometric members.}
\label{tab:parallax}
\begin{tabular}{cccccccccc}
\hline
\hline
 ID Gaia & ID 2MASS & $\mu_{\alpha}\,cos \delta$ & $\mu_{\delta}$ & $G$ & $\pi$ & $\sigma{_\pi}$ & $f$ & $rgeo$ &  $rpgeo$ \\
5878925.... & J144.... & [mas\,yr$^{-1}$] & [mas\,yr$^{-1}$] & [mag] & [mas] & [mas] & & [pc] & [pc] \\
\hline
503633099264  & 50584-5949317 & -4.745 & -2.115 & 18.5 & 0.0501 &	0.2334 & 4.66 & 4005 &		4409 	\\
709791531776  & 50217-5949281 & -4.799 & -2.714 & 19.6 & 0.6851 &	0.3978 & 0.58 & 3221 &	5343	\\
503620821120  & 50498-5949434 & -4.759 & -2.148 & 18.2 & 0.4277 &	0.1920 & 0.45 & 3159 &		3107	\\
503620808704  & 50973-5949588 & -5.201 & -3.611 & 18.2 & 0.4047 &	0.1798 & 0.44 & 4993 &		2342 \\
499327763456  & 50647-5950131 & -4.661 & -2.698 & 18.6 & 0.5566 &	0.2109 & 0.38 & 3033 &		3219 	\\
709791805824  & 50320-5949440 & -4.927 & -2.627 & 16.9 & 0.2405 &	0.0866 & 0.36 & 3567 &	3527\\
709791532928  &  50240-5949199 & -4.675 & -2.626 & 17.9 & 0.3854 &	0.1353 & 0.35 & 2863 &		2766\\ 
705486190208  & 50460-5949095 & -4.909 & -2.216 & 19.0 & 0.9958 &	0.2674 & 0.27 & 1223 &		3011 	\\
709779257984  &  50222-5949405 & -5.220 & -2.454 & 17.5 & 0.4569	& 0.1196 & 0.26 & 2103 & 2095	\\
\hline
\end{tabular}
\parbox{16cm}{
{Notes: 
{\it Gaia} EDR3 and 2MASS are identifications, 
$\mu_{\alpha} cos \delta$ and $\mu_{\delta}$ the proper motions, 
$G$ magnitude, 
$\pi$ stellar parallax, $\sigma{_\pi}$ error of the stellar parallax, $f$ parallax divided by its standard error, 
$rgeo$ median of the geometric distance posterior and $rpgeo$ median of the photogeometric distance posterior explained in Sect.~\ref{sec:dist} \citep{bai21}.} }
\end{center}
\end{table*}

\end{appendix}

\end{document}